\documentclass[letterpaper,10pt]{article}

\usepackage[letterpaper,margin=1.15in]{geometry}
\usepackage{authblk}

\usepackage{caption}
\usepackage{subcaption}
\usepackage{tcolorbox}
\usepackage{makecell}
\usepackage{booktabs}
\usepackage{tabularx}
\usepackage{multirow}
\usepackage{graphicx}
\usepackage{comment}
\usepackage{enumitem}
\usepackage{tikz}
\usepackage{amsmath}
\usepackage{amssymb}
\usepackage{amsthm}
\usepackage{algorithm}
\usepackage{algpseudocode}
\usepackage{xcolor}
\usepackage{enumitem}
\usepackage{hyperref}
\hypersetup{colorlinks=true,allcolors=blue}
\usepackage{tikz}
\usetikzlibrary{arrows.meta, positioning,calc,backgrounds,fit}

\definecolor{true11}{RGB}{8,69,148}    
\definecolor{cold}{RGB}{66,146,198}    
\definecolor{binc}{RGB}{110,110,110}   
\definecolor{ann}{RGB}{60,60,60}       
\definecolor{symc}{RGB}{25,25,25}      
\definecolor{obsc}{RGB}{90,122,153}    
\definecolor{alicefill}{RGB}{222,235,247}
\definecolor{bobfill}{RGB}{224,236,222}
\definecolor{evefill}{RGB}{250,232,232}
\definecolor{evestroke}{RGB}{178,54,54}
\definecolor{chanfill}{RGB}{242,242,238}
\definecolor{keycol}{RGB}{150,115,20}
\definecolor{annc}{RGB}{60,60,60}
\tikzset{
  binline/.style={draw=binc!80, line width=0.8pt, dash pattern=on 2pt off 1.5pt},
  symtick/.style={draw=symc, line width=1.2pt},
  obstick/.style={draw=obsc, line width=1.2pt, dash pattern=on 1.5pt off 1pt}}

\newtheorem{theorem}{Theorem}
\newtheorem{definition}{Definition}

\newtheorem{remark}{Remark}

\definecolor{spkA}{HTML}{2F5D8A}   
\definecolor{spkB}{HTML}{3F7A5A}   
\definecolor{turnbg}{HTML}{F4F4F2} 
\definecolor{metafg}{HTML}{6B6B6B} 

\newcommand{\meta}[3]{%
  {\footnotesize\color{metafg}\textsf{#1 \,|\, #2 \,|\, #3}}%
}

\newcommand{\turn}[5]{%
  \par\vspace{2pt}%
  \begin{flushleft}
  \end{flushleft}%
}

\newcommand{\turnA}[2]{%
  \par\vspace{3pt}\noindent
  \begin{minipage}{0.72\linewidth}
    \colorbox{turnbg}{\parbox{\dimexpr\linewidth-2\fboxsep}{%
      {\footnotesize\color{spkA}\textbf{\textsf{Agent A}}}\\[1pt]
      \small #1\\[2pt]#2}}
  \end{minipage}\hfill\mbox{}\par
}

\newcommand{\turnB}[2]{%
  \par\vspace{3pt}\mbox{}\hfill
  \begin{minipage}{0.72\linewidth}
    \colorbox{turnbg}{\parbox{\dimexpr\linewidth-2\fboxsep}{%
      \raggedleft
      {\footnotesize\color{spkB}\textbf{\textsf{Agent B}}}\\[1pt]
      \small\raggedleft #1\\[2pt]#2}}
  \end{minipage}\par
}

\begin{document}

\date{}

\title{\Large \bf Feedback Coding Enables Inference-Time Covert Agentic Communication}

\author[1]{Sidong Guo}
\author[2]{Sajani Vithana}
\author[3]{Atefeh Gilani}
\author[3]{Lalitha Sankar}
\author[3]{Oliver Kosut}
\author[2]{Flavio P. Calmon}
\affil[1]{Georgia Institute of Technology}
\affil[2]{Harvard University}
\affil[3]{Arizona State University}

\maketitle

\subsection*{Abstract}
As large language models (LLMs) are increasingly used to automate digital interactions, users can leverage LLM-generated text as cover for covert communication within seemingly benign conversations. Existing LLM steganography, however, is predominantly white-box, requiring the sender and receiver to share the cover statistics, typically through access to the model weights and prompt. Black-box schemes remove this requirement by allowing the receiver to operate solely on the generated text, but current approaches rely on fixed-length, open-loop watermarking techniques that suffer from high decoding error rates under variable-length token generation. We recast black-box LLM steganography as a sequential communication problem with causal, noiseless feedback: every generated token is observed by both parties and can guide subsequent embedding. Based on this perspective, we introduce \textbf{B}urnashev \textbf{A}daptive Posterior \textbf{M}atching (BAM), a feedback-coding scheme that combines posterior matching with a decode-and-confirm phase. The design is inspired by classical information-theoretic feedback-coding principles, while its security is established through a cryptographic reduction proof. Across three open-weight language models, we demonstrate that BAM attains 0-0.1\% empirical message error on an 8-bit payload in around 50 tokens, across 1000 trials, versus 10-17\% for the strongest black-box baseline at comparable length. Building on the proposed steganography algorithm, we demonstrate the feasibility of an end-to-end communication protocol that achieves high communication rates across multiple conversational settings. 

\section{Introduction}
Steganography is the practice of embedding a secret payload into an innocuous covertext such that only an intended receiver can reliably recover the hidden message while an adversary cannot distinguish the resulting stegotext from an ordinary cover. Steganographic communication has been studied extensively in both the information-theoretic literature \cite{bloch2016covert} and the cryptography literature \cite{fridrich2009steganography}. Its asymptotic embedding capacity, under a wide range of cover distributions and distortion constraints, has been well characterized \cite{ker2008square,filler2009square,moulin2003information}. 

The success of generative models, particularly large language models (LLMs), has renewed interest in text steganography, where secret messages are embedded directly into LLM-generated text. One emerging application of generative steganography is the establishment of covert communication channels mediated by LLM agents. Users can leverage LLM systems as a cover to exchange hidden messages beneath an overt conversation that remains fluent and task-consistent, effectively evading surface-level content policing. This threat model encompasses multiple configurations: two humans using agents as proxies in AI-mediated conversations or a human coordinating with a proxy agent \cite{hancock2020ai, mieczkowski2021ai,jones2026turing}. In these settings, human actors may leverage local control over the agent's sampling loop to apply logit-steering algorithms, embedding secret messages directly into covertext to evade surface-level monitoring. This capability is becoming increasingly realistic, as suitable models can sustain dialogues sufficiently natural to pass a standard three-party Turing test \cite{jones2026turing}.

Realizing this threat, however, requires operating under strict practical constraints. Existing steganography approaches can be broadly divided according to the information available to the message decoder. The dominant paradigm is \textit{white-box} steganography, in which both encoder and decoder have access to the language model's conditional next-token distribution, which often entails sharing model parameters and
prompts. Representative methods include coupling-based approaches that construct a minimum-entropy coupling between the message and cover
distributions \cite{de2022perfectly},
sampling schemes that mask the message with pseudorandom bits and embed it through inverse-transform sampling of the model distribution
\cite{kaptchuk2021meteor}, and arithmetic-coding-based approaches \cite{ziegler2019neural}. Despite their algorithmic differences, these
techniques require the sender and receiver to share the same cover
statistics, an assumption that rarely holds for independently deployed LLM agents.

Consequently, realistic covert communication requires a \textit{black-box} approach \cite{zamir2024undetectable}\cite{ryabko2009asymptotically}, where the receiver observes only the generated text and has no access to the underlying language model weights, prompt, or token probabilities. Yet, existing black-box techniques remain ill-suited for conversational steganography. Most recent works in this domain take the form of inference-time multi-bit watermarking \cite{gilani2026arcmark,yoo2023advancing,feng2025bimark,cui2026mc}. Viewed as communication systems, these methods are largely fixed-length and open-loop: the encoder converts messages to a fixed set of codewords without using the realized outputs to adapt future transmissions, and the decoder commits after a preset token budget. Consequently, fixed-length protocols fail in dynamic conversational settings due to two main limitations:
\begin{itemize}
    \item \textbf{High Decoding Errors at High Embedding Rates:} Existing schemes can only reliably recover payloads at very low embedding rates. Attempting higher transmission rates dramatically increases decoding error, falling well short of the error-rate trade-offs achieved by white-box alternatives.
    \item \textbf{Inflexibility to Variable Turn Lengths:} Natural conversational turns vary unpredictably in length. Fixed-length protocols suffer decoding failures when a turn terminates prematurely, and their fixed codebooks cannot efficiently adapt when a turn naturally extends beyond its expected budget.
\end{itemize}

To address these limitations, we develop a provably secure \textit{black-box} steganographic scheme based on sequential feedback communication. The use of feedback in black-box steganography was previously explored by Zamir~\cite{zamir2024undetectable}, who combines repeated single-symbol embedding with a feedback-based error-correction mechanism to enable reliable multi-bit transmission. We construct a fully operational scheme by instead take an information-theoretic feedback-coding approach in which feedback directly governs the evolution of the code itself. Since every generated token is observed identically by both communicating agents, the encoder can reproduce the decoder's state and adapt future codeword symbols to the realized transcript. Specifically, we construct a variable-length protocol that combines adaptive posterior matching~\cite{shayevitz2011optimal} with a Yamamoto-Itoh confirmation mechanism~\cite{yamamoto1979asymptotic}. Rather than encoding and correcting individual payload symbols separately, the protocol maintains a belief over the entire message space, adaptively selects subsequent channel inputs from this belief, and verifies tentative decisions through an explicit confirmation phase. This enables the token budget to adapt to the realized channel while achieving high rate efficiency and near-zero decoding error without requiring the decoder to access the underlying language model. We further extend this point-to-point protocol to conversational settings and demonstrate the feasibility of covert agentic communication in terms of reliability, rate, and quality.

\paragraph{Contributions.}
\begin{itemize}[leftmargin=1.2em,itemsep=2pt,topsep=2pt]
\item We demonstrate the feasibility of inference-time covert agentic communication by modeling the black-box steganographic channel as a communication channel with causal, noiseless feedback, providing a formal system model and empirical frameworks for evaluating this emerging threat (Section~\ref{sec:threat}).

\item We introduce Burnashev Adaptive Posterior Matching (BAM), a variable-length feedback coding scheme explicitly designed for the black-box setting. BAM combines a sequential probability-matching step to localize messages with a confirmation phase to verify tentative decisions. This architecture substantially improves reliability over existing baseline methods (Section~\ref{sec:method}).

\item We provide a cryptographic reduction proving computational indistinguishability of the proposed system under pseudo-random function (PRF) security (Section~\ref{sec:security}).

\item We develop and implement a practical conversational protocol that integrates BAM with natural turn-taking, early-stopping generation, and multi-turn payload transmission (Section~\ref{sec:convo}).

\item We conduct extensive evaluations across multiple language models, moving beyond standard multi-bit watermarking primitives to test interactive conversational settings. Our results demonstrate substantial improvements in embedding efficiency and decoding reliability (Section~\ref{sec:experiment}).
\end{itemize}

Taken together, this work establishes that efficient inference-time covert communication is highly feasible for conversational settings, even under prompt and model agnostic settings. 

\section{Background and Related Work}\label{sec:background}
\paragraph{Information-Theoretic and Cryptographic Security Measures}
In this work, we evaluate our steganographic framework under both information-theoretic security adapted from Cachin~\cite{cachin2004information} and cryptographic security. Information-theoretic security traditionally relies on notions of \textit{weak} and \textit{strong secrecy} to guarantee that the information leakage of a message is either vanishing on average or can be made arbitrarily small for unnormalized mutual information \cite{wyner1975wire, bloch2013strong}. Information-theoretic \textit{covertness} \cite{bloch2016covert}, on the other hand, requires statistical indistinguishability between covertext and stegotext. Because these measures assume a uniform message prior, modern frameworks combine covertness and strong secrecy by requiring the Kullback-Leibler divergence between the cover and stego distributions to be arbitrarily small for every message \cite{cachin2004information}. While Cachin's framework establishes theoretical secrecy against a passive eavesdropper, it does not guard against an active adversary with API access. Consequently, provably secure steganographic frameworks also require \textit{distinguishing security} \cite{pang2025provable}, which ensures computational indistinguishability against any probabilistic polynomial-time adversary. 

\paragraph{Inference-Time Multi-Bit Watermarking}
As opposed to training-time steganography, which embeds information during the training process, 
inference-time steganography often induces a coupling between cover and stegotext to ensure indistinguishability \cite{de2022perfectly}. 
The core of our distortion-free guarantee builds on the optimal transport (OT) coupling of ArcMark \cite{gilani2026arcmark}. Conceptually, ArcMark is an inference-time multi-bit watermarking technique that embeds information by projecting both the language model's vocabulary and the hidden message symbols onto a unit circle. 
To transmit a message, the encoder biases the next token distribution to favor the selection of tokens that are positioned angularly closer to the target message symbol on the circle. To ensure the quality of generation, ArcMark formulates the channel synthesis process as an OT problem with distortion-free constraint. Solving this optimization minimizes the expected circular distance between the sampled tokens and the target messages, subject to a marginal constraint.

\paragraph{Information-Theoretic Feedback Coding}
By recognizing that autoregressive token generation naturally forms a channel with instantaneous feedback, our scheme adapts classical feedback principle, combining posterior matching \cite{shayevitz2011optimal} with a Yamamoto-Itoh-style confirmation phase~\cite{yamamoto1979asymptotic}. Classical information-theoretic results establish that noiseless feedback, while not increasing the capacity of a memoryless channel \cite{cover1999elements}, can dramatically improve reliability function (the rate at which decoding error decreases) through adaptive coding. Foundational results include Burnashev's optimal reliability exponent \cite{burnashev1976data}, Yamamoto-Itoh's two-phase achievability strategy \cite{yamamoto1979asymptotic} and modern variable-length feedback coding schemes \cite{polyanskiy2011feedback}. In addition to optimal exponent, posterior matching \cite{shayevitz2011optimal} provides a capacity-achieving sequential coding strategy in which the encoder selects channel inputs according to the decoder's current posterior belief.

\paragraph{Covert Agentic Communication}
Our work is orthogonal to existing literature that focuses on eliciting, detecting, or evaluating the impact of deceptive agentic behaviors \cite{motwani2024secret,xu2025comet,huang2026whispering}. We approach this threat as a sequential communication and cryptographic security problem at inference time. \textit{Covert Agentic Communication} therefore refers to the ability of LLM agent (either independent or prompted by human actors) to hide messages in generated texts in inference-time. Covert communication has emerged as a significant security concern within LLM-based multi-agent systems and AI-mediated interactions \cite{hancock2020ai, mieczkowski2021ai}, where \cite{vaikuntanathan2026undetectable} showed that covert communication between agents are possible in the absence of shared secret via cryptographic key exchange.

\section{System Framework and Threat Model}\label{sec:threat}

\subsection{Notations and Problem Setup}
We use the following notations throughout the paper: 
\begin{itemize}
   \item \textbf{Variables.}
    Random variables and their realizations are denoted by uppercase and lowercase letters, respectively (e.g., \(X\) and \(x\)). Let \(X\) take values in a finite alphabet \(\mathcal X\). For \(n\in\mathbb N\), we write \(X^n\triangleq(X_1,\ldots,X_n)\in\mathcal X^n\), and for \(1\le i\le j\le n\), we denote the subsequence by \(X_{i:j}\triangleq(X_i,\ldots,X_j)\). 

    \item \textbf{Parameters.}
    Let \(\mathcal X\) denote the token vocabulary with cardinality \(|\mathcal X|=V\). We denote by \(\mathcal S\subseteq\Delta^V\) the set of possible next-token distributions, where \(\Delta^V\) is the probability simplex over $V$ tokens. At token position \(t\), the language model induces a distribution \(S_t=s_t\in\mathcal S, S_t\triangleq P_{X_t|X_{1:t-1},\rho}\), determined by the model parameters, prompt \(\rho\), and previously generated tokens \(x^{t-1}\) (when clear from context, we suppress the dependence on \(\rho\) in notation). A binary message of length \(\ell\) is denoted by \(m\in\mathcal M\triangleq\{0,1\}^{\ell}\), we use idx($m$) to denote the corresponding message index (decimal representation). Each message is encoded into a codeword \(U^n(m)\in\mathcal U^n\) (may be stochastic), where \(n\) denotes the (possibly variable) codeword length. We use $\tau$ to denote the length of generated text, where we require $n\leq \tau$. We also use \(\kappa\) to denote shared seed and \(K=k\in \mathcal{K}\) as generated keys. 
    \item \textbf{Metrics.}
    The Kullback--Leibler (KL) divergence between probability mass functions $P_1$ and $P_2$ on $\mathcal X$ is
    \begin{equation}
        \mathbb D(P_1\Vert P_2)
        \triangleq
        \sum_{x\in\mathcal X}P_1(x)\log\frac{P_1(x)}{P_2(x)}.
    \end{equation}
    Logarithms are taken to base $2$.
    We use \(\lambda\) to denote the security parameter and \(1^\lambda\) its unary representation. A function \(\operatorname{negl}(\lambda)\) is negligible if it vanishes faster than the reciprocal of every polynomial in \(\lambda\). The communicating parties share a uniform secret seed $\kappa \in \{0,1\}^\lambda$. An API adversary is denoted by $\mathcal{A}$. The binary decision event $\mathcal{A}(1^\lambda) = 1$ indicates that the adversary classifies the output as stegotext, whereas $\mathcal{A}(1^\lambda) = 0$ indicates it is classified as benign covertext.
 \end{itemize}
\subsection{System Model}
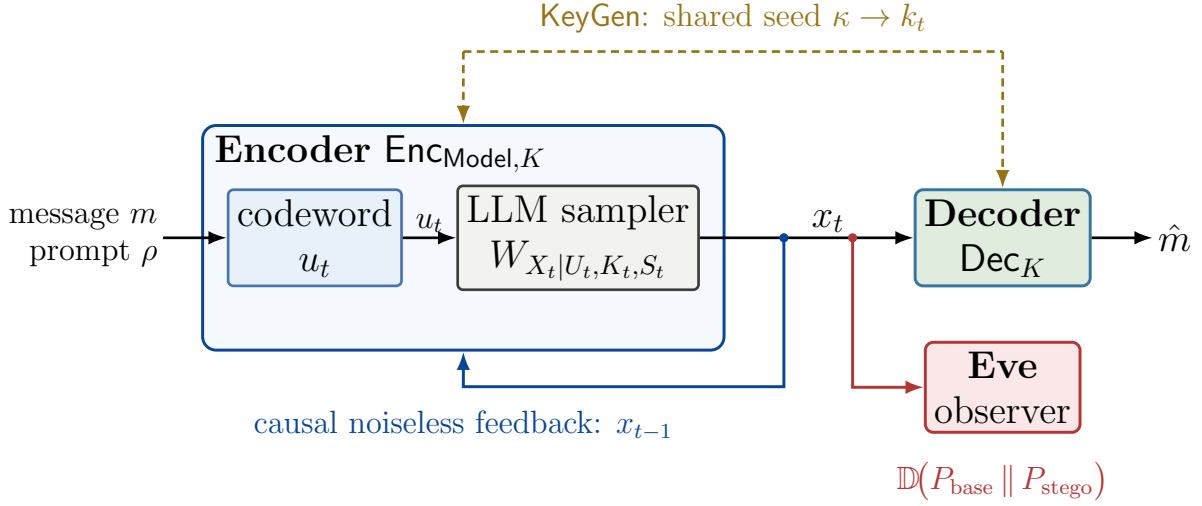
\begin{figure}[t]
  \centering
  \resizebox{\columnwidth}{!}{%
  \begin{tikzpicture}[
    font=\Large,
    line width=1.0pt,
    >={Latex[length=2.4mm]},
    proc/.style={draw, rounded corners=2pt, align=center, line width=1.0pt,
                 minimum width=1.8cm, minimum height=0.95cm, fill=white},
    party/.style={draw, rounded corners=3pt, align=center, line width=1.1pt,
                  minimum width=2.0cm, minimum height=1.1cm},
    arr/.style={-{Latex[length=2.4mm]}, line width=1.0pt},
    lab/.style={font=\normalsize, inner sep=1.5pt},
  ]
    \node[proc, fill=alicefill!70, draw=true11!75] (cw)
          {codeword\\[-1pt]$u_t$};
    \node[proc, right=0.7cm of cw, fill=chanfill, draw=annc,
          minimum width=2.2cm] (samp)
          {LLM sampler\\[-1pt]$W_{X_t\mid U_t,K_t,S_t}$};
    \draw[arr] (cw) -- node[lab,above,pos=0.5,font=\large]{$u_t$} (samp);
    \begin{scope}[on background layer]
      \node[draw=true11, line width=1.1pt, rounded corners=5pt, fill=alicefill!30,
            fit=(cw)(samp), inner xsep=9pt, inner ysep=22pt] (enc) {};
    \end{scope}
    \node[lab, anchor=north west, text=black, font=\Large] at ($(enc.north west)+(3pt,-2pt)$)
      {\textbf{Encoder} $\mathsf{Enc}_{\mathsf{Model},K}$};
    \node[lab, left=0.5cm of enc.west, align=right, anchor=east, font=\large] (msg)
      {message $m$\\ prompt $\rho$};
    \draw[arr] (msg) -- (cw.west);
    \node[party, right=2.8cm of samp, fill=bobfill, draw=cold!80!black] (dec)
          {\textbf{Decoder}\\[-1pt]\Large $\mathsf{Dec}_{K}$};
    \node[lab, right=0.8cm of dec, font=\Large] (mhat) {$\hat m$};
    \coordinate (tap) at ($(samp.east)!0.5!(dec.west)$);
    \draw[arr] (samp.east) -- node[lab,above,pos=0.6,font=\Large]{$x_t$} (dec.west);
    \draw[arr] (dec.east) -- (mhat);
    \node[party, draw=evestroke, fill=evefill, below=0.7cm of dec] (eve)
          {\textbf{Eve}\\[-1pt]\Large observer};
    \coordinate (tapL) at ($(tap)+(-0.30,0)$);
    \coordinate (tapR) at ($(tap)+(0.60,0)$);
    \coordinate (rail) at ($(tap)+(0,-1.75)$);
    \coordinate (railL) at (tapL |- rail);
    \coordinate (railR) at (tapR |- rail);
    \fill[evestroke] (tapR) circle (1.7pt);
    \draw[arr, draw=evestroke, line width=1.1pt]
      (tapR) |- (eve.west);
    
    \node[lab, below=0.3cm of eve, align=center, text=evestroke, font=\large] (div)
          {$\mathbb{D}\!\big(P_{\mathrm{base}}\,\|\,P_{\mathrm{stego}}\big)$};
    \coordinate (kL) at (enc.north);
    \coordinate (kR) at (dec.north);
    \coordinate (kmid) at ($(kL)!0.5!(kR)$);
    \node[lab, text=keycol, anchor=south, font=\large] (key) at ($(kmid)+(0,1.55)$)
          {$\mathsf{KeyGen}$: shared seed $\kappa \to k_t$};
    \coordinate (krail) at ($(key.south)+(0,-0.18)$);
    \coordinate (kLr) at (kL |- krail);
    \coordinate (kRr) at (kR |- krail);
    \draw[keycol, dash pattern=on 3pt off 2pt] (kLr) -- (kRr);
    \draw[arr, draw=keycol, dash pattern=on 3pt off 2pt] (kLr) -- (enc.north);
    \draw[arr, draw=keycol, dash pattern=on 3pt off 2pt] (kRr) -- (dec.north);
    \coordinate (railfb) at ($(tap)+(0,-1.95)$);
    \coordinate (railLfb) at (tapL |- railfb);
    \coordinate (encfb) at ($(enc.south)+(-0,0)$);
    \fill[true11] (tapL) circle (1.7pt);
    \draw[arr, draw=true11, line width=1.1pt]
      (tapL) -- (railLfb) -| (encfb);
    \node[lab, text=true11, anchor=north, font=\large] at ($(enc.south)+(0,-0.7)$)
      {causal noiseless feedback: $x_{t-1}$};
  \end{tikzpicture}}
  \caption{System model for a single generated token at position $t$.}
  \label{fig:system}
\end{figure}

The overall framework of the covert communication setup is depicted in Fig.~\ref{fig:system}, which illustrates a single token generation at token position $t$. An encoder receives the prompt and the secret message, which, together with the accumulated history so far, determine the next-token distribution and the codeword symbol $u_t$, respectively. The system then samples a token $x_t$ from either the biased distribution or the clean distribution, which is subsequently observed by both the legitimate decoder and an external observer (\emph{Eve}). We define the core LLM communication channel as follows:

\begin{definition}[LLM channel] \label{channel}
An LLM channel \(\mathsf{Model}\) is a probabilistic sampling rule that, on input prompt \(\rho\) and token history \(x^{t-1}\in\mathcal X^{t-1}\), induces the next-token distribution \(s_t\). Without hidden message embedding, it samples
\begin{equation}
    s_t\leftarrow\mathsf{Model}(\rho,x^{t-1}),
    \qquad
    x_t\sim W_{X|S}(\cdot|s_t),
\end{equation}
where \(W_{X|S}(x_t|s_t)\) is the unbiased base sampling rule (e.g. Top-k with temperature $1.0$).
Given an embedding symbol \(u_t\in\mathcal U\) and key \(k_t\in\mathcal K\), the steganographic sampler instead generates
\begin{equation}
    x_t\sim W_{X|U,K,S}(\cdot|u_t,k_t,s_t).
\end{equation}
\end{definition}

\begin{definition}[Black-box steganographic scheme]\label{stegosystem}
A black-box steganographic scheme for an LLM channel $\mathsf{Model}$ is a tuple of probabilistic algorithms
\begin{equation}
    \Pi_{\mathsf{Model}}
    =  \left(\mathsf{KeyGen},\mathsf{Enc}_{\mathsf{Model},K},\mathsf{Dec}_K\right).
\end{equation}
The algorithms are defined as follows.
\begin{itemize}
    \item $\mathsf{KeyGen}(1^\lambda)$ takes a shared seed $\kappa\in\{0,1\}^{\lambda}$ and outputs the secret keying material $K$ used by the encoder and decoder.
    \item $\mathsf{Enc}_{\mathsf{Model},K}(m,\rho)\rightarrow x^\tau$ takes a message $m\in\mathcal M$ and prompt $\rho$, and outputs a variable-length stegotext $x^\tau$. It has inference-time access to $\mathsf{Model}$ and its next-token distribution. 
    \item $\mathsf{Dec}_{K}(x^\tau)\rightarrow\hat m$ takes the stegotext and outputs a message estimate. It does not require access to $\mathsf{Model}$, the prompt, or token probabilities.
\end{itemize}
\end{definition}

\begin{remark}[Noiseless feedback]
    In the absence of editing, the dependence on \(X^{t-1}\) models perfect causal feedback: after each token is generated, both communicating parties observe the same transcript and can derive identical side information. This feedback is the source of the reliability gains achieved by our algorithm. 
\end{remark} 
\begin{remark}[System assumptions]
    Specifically, we refer to \emph{white-box} steganography as schemes that require decoder access to cover statistics \cite{liao2025framework}. In LLM steganography, this means access to model weights and prompts. Unlike \emph{white-box} steganographic schemes, in our setting the decoder is agnostic to the sender's model weights, prompt, and token probabilities. Consequently, synchronization of side information is achieved entirely through deterministic hashing of the shared transcript \cite[Appendix~A]{tsur2026heavywater}. However, based on the assumption of Definition ~\ref{stegosystem}, apart from the public algorithm, decoder knows the sender tokenizer and vocabulary $\mathcal{X}$. The encoder, in contrast, has inference-time access to the underlying language model, the prompt used, and therefore its next-token distribution \(s_t\). An information access table for our proposed algorithm is provided in Appendix~\ref{app:party-info}.
\end{remark}

\subsection{Adversarial Model}
As illustrated in Fig.~\ref{fig:system}, the system models potential adversary at two intervention points: the input side, where the prompt and message are chosen, and the observation side (Eve), where the public transcript is inspected. Based on these roles, we define two distinct adversarial settings:

\begin{enumerate}[leftmargin=1.2em]
    \item \textbf{Passive Eavesdropper}: In this setting, legitimate party chooses the prompt and message, while Eve is an adversary. This models an entity such as a platform content moderator or an automated firewall passively analyzing the generated text in transit. The passive adversary's primary objective is to detect whether the transcript contains a hidden payload (\textit{covertness}). Alternatively, Eve may attempt to decode the payload if its existence is suspected (\textit{secrecy}). Security against this adversary is evaluated using information-theoretic divergence strengthened from \cite{cachin2004information}, which guarantees both \textit{covertness} and \textit{secrecy}.
    
    \item \textbf{Active API Adversary}: In this setting, the adversary controls both the input side and the observation side. This models an auditor or malicious user who adaptively queries the deployed language model via its API, providing chosen prompts and message to elicit specific outputs. Operating under a probabilistic polynomial-time constraint, the adversary attempts to distinguish between generated stegotext and covertext. Security against this adversary is evaluated using cryptographic distinguishing advantage. In this paper, we consider a stronger version of the API adversary that can query both message and prompt. 
\end{enumerate}
Since our primary application is LLM steganography, where stegotext is often assumed to be shared, an adversary who edits, deletes, or paraphrases the transcript is outside the scope of our security assumptions.

\subsection{Security and Performance Definitions}

\begin{definition}[$\delta$-Reliability]\label{def:reliability}
    A black-box steganographic scheme is $\delta$-reliable if there exists \(\delta \in [0,1]\) such that 
    \begin{equation} \label{reliability}
        P_{\mathrm e}=\frac{1}{|\mathcal{M}|}\sum_{m=1}^{|\mathcal{M}|}\Pr\left[\mathsf{Dec}_{K}(X^\tau)\neq m|M=m\right] \leq \delta . 
    \end{equation}
    In addition, we define the finite-length reliability function as 
    \begin{equation} \label{reliability function}
        E \triangleq -\frac{1}{\mathbb{E}[\tau]} \log P_{\mathrm e}.
    \end{equation}
\end{definition}

\begin{definition}[Embedding rate]\label{rate}
For a message set $\mathcal M$ and variable transcript length $\tau$, the embedding rate is
\begin{equation}
    R
    \triangleq
    \frac{\log|\mathcal M|}{\mathbb E[\tau]}.
\end{equation}
\end{definition}

\begin{definition}[$\sigma$-secure on average (against passive adversary)]\label{cachin}
    A black-box steganographic scheme is $\sigma$-secure on average if for any channel input \(u\) (Definition ~\ref{channel}) and state \(s\)
    \begin{equation}
        \sup_{s,u}\mathbb{D}\bigg(W_{X|S}(x|s)||\sum_{k\in \mathcal{K}}P_{K}(k)W_{X|U,K,S}(x|u,k,s)\bigg) \leq \sigma,
    \end{equation}
    where \(P_{K}(k)\) is the induced key distribution. When $\sigma = 0$, the scheme is said to be perfectly secure on average. 
\end{definition}

\begin{definition}[Cryptographic security (against polynomial-time adversaries)]\label{crypto}
A black-box steganographic scheme is cryptographically secure if, for any polynomial-time API adversary \(\mathcal{A}\) making polynomially many adaptive queries, the following advantage is negligible:
\begin{equation}\label{advantage}
    \left|
    \Pr\!\left[\mathcal{A}^{\mathsf{Real}}(1^\lambda)=1\right]
    -
    \Pr\!\left[\mathcal{A}^{\mathsf{Clean}}(1^\lambda)=1\right]
    \right|
    \leq \operatorname{negl}(\lambda).
\end{equation}
On query \(i\), \(\mathcal A\) adaptively chooses a prompt-message pair \((\rho_i,m_i)\). In the \(\mathsf{Real}\) experiment, the oracle returns stegotext generated by \(\Pi_{\mathsf{Model}}\) from \((\rho_i,m_i)\); in the \(\mathsf{Clean}\) experiment, the oracle returns covertext generated by the base channel \(W_{X|S}\) from the same prompt \(\rho_i\), independently of \(m_i\).
\end{definition}

\section{Feedback Coding and the BAM Protocol}\label{sec:method}
We present the covert communication protocol in two stages that together form a single variable-length feedback code. In the first stage, variable-length posterior matching (Section~\ref{sec:pm}), progressively localizes the hidden message by steering the decoder's posterior belief. Once the decoder's confidence exceeds a prescribed threshold, a second confirmation phase (Section~\ref{sec:bam}) verifies the tentative decision and amplifies decoding reliability. We refer to the resulting feedback code as \emph{Burnashev Adaptive Posterior Matching (BAM)}. The name reflects BAM’s architectural inspiration from the Yamamoto--Itoh strategy \cite{yamamoto1979asymptotic}, which achieves the \textit{Burnashev reliability function} for classical memoryless channels \cite{burnashev1976data}. We do not claim the corresponding asymptotic optimality for the LLM channel studied here; instead, we empirically evaluate the finite-length gains contributed by variable-length stopping and confirmation.

\subsection{Intuition}

As discussed in Definition~\ref{channel}, LLM steganography naturally induces a communication channel with causal, noiseless feedback. From an information-theoretic perspective, feedback does not increase the asymptotic capacity of a memoryless channel \cite{cover1999elements}. Its principal benefit is instead to improve the \emph{reliability function}(\ref{reliability function}), which characterizes the exponential decay of the decoding error probability with expected blocklength \cite{burnashev1976data,yamamoto1979asymptotic,polyanskiy2011feedback}. Consequently, we will show that embedding rate alone does not fully characterize the performance of a steganographic scheme; practical covert agentic communication equally depends on how reliably messages can be recovered within a finite number of tokens available.

Two distinct mechanisms underlie the reliability gain \cite{naghshvar2012} (increase in reliability function). The first is \emph{sequentiality gain}, whereby the transmitter communicates using a variable stopping time, spending additional channel uses when the decoder remains uncertain while terminating early once sufficient confidence has been established. The second is \emph{adaptivity gain}, whereby future channel inputs are selected according to past observations. To this end, our algorithm incorporates three core mechanism previously unexplored in black-box steganography:
\begin{itemize}
    \item \textbf{Posterior matching:} An encoding technique by which encoder draws codeword symbols based on decoder's message belief \cite{shayevitz2011optimal}. We adopted a mismatched-belief variation to enable efficient, higher-rate encoding, while eliminating the need to store an exponentially large pre-shared codebook.
    \item \textbf{Threshold decoding:} We achieve \emph{sequentiality gain} by using a threshold decoder based on posterior belief, which enables variable-length stopping.
    \item \textbf{Yamamoto-Itoh scheme:} Yamamoto-Itoh's classical result shows that \emph{adaptivity gain} can be exploited by segment-based adaptation through a short confirmation phase, yielding the optimal reliability exponent (Burnashev exponent) for communication with feedback \cite{yamamoto1979asymptotic,polyanskiy2011feedback}.
\end{itemize}
An empirical evaluation of these performance gains is provided in Appendix \ref{app:extra}.

\subsection{Feedback Coding via Posterior Matching}\label{sec:pm}

At a high level, this point-to-point communication method operates through four sequential components:
\begin{enumerate}
    \item \textbf{Codebook generation:} The encoder dynamically maps the intended message to a channel input based on the decoder's current posterior belief (Codeword in Fig. \ref{fig:system}).
    \item \textbf{Key generation and optimal transport:} Both parties derive synchronized cryptographic side information from the shared transcript to define a secure, distortion-free coupling using optimal transport (LLM sampler in Fig. \ref{fig:system}).
    \item \textbf{Belief computation and encoding:} The decoder (and the synchronized encoder) updates its posterior distribution over the message space using a robust likelihood model (Encoder in Fig. \ref{fig:system}).
    \item \textbf{Decoding:} The decoder applies a threshold-based rule to the posterior to commit to the transmitted message (Decoder in Fig. \ref{fig:system}).
\end{enumerate}

\subsubsection{Codebook Generation}

Let \(m \in \mathcal{M}\) denote the true message, and let \(\pi_t(m)\) denote the decoder's posterior belief \footnote{with a slight abuse of terminology, here decoder's belief is not the true Bayesian belief since decoder has no access to the channel.} on message \(m\) after observing the first \(t\) generated tokens. We initialize the decoder with the uniform prior \(\pi_{0}(m')=\tfrac{1}{|\mathcal{M}|},\ \forall m' \in \mathcal{M}\). 

To select the next channel input, the encoder maps the discrete message space onto the continuous unit interval \([0, 1]\) by partitioning it into sub-intervals proportional to each message's current posterior probability. Given the posterior distribution \(\{\pi_{t-1}(m')\}_{m'\in \mathcal{M}}\), the encoder determines the input codeword symbol at token position \(t\) according to the posterior matching rule $u_{t} = F^{-1}_{U}(V_{t}(m))$, 
\begin{equation}\label{codeword generation}
    V_{t}(m) = \sum_{\substack{m': \textnormal{idx}(m') < \textnormal{idx}(m)}} \pi_{t-1}(m') + R_{t}\pi_{t-1}(m).
\end{equation}
\(V_t(m)\) represents a continuous cumulative belief statistic for message \(m\) (the sum of belief of messages with message index lower than $m$). The summation term calculates the lower bound of message \(m\)'s probability interval. The term \(R_t\), idealized as \(R_t\sim\textnormal{Unif}[0,1]\), is synchronized pseudorandomness derived from the shared seed, which we explain in subsequent sections. This randomized mapping ensures that \(V_t\) remains continuously distributed on \([0, 1]\). Finally, for the codeword random variable $U$ with cardinality $|\mathcal{U}|=p$, \(F^{-1}_{U}\) denotes inverse-CDF sampling, which transforms this continuous belief statistic into a discrete codeword symbol. In the classical posterior matching framework, the input distribution \(F_U\) is chosen to maximize the achievable rate of the communication channel \cite{shayevitz2011optimal}. Since the effective LLM channel depends on the prompt and generally does not admit a tractable capacity characterization, we instead choose \(F_U\) to be uniform over \(\mathcal U\), yielding \(u_t=\lfloor pV_t(m)\rfloor\), where \(p=|\mathcal{U}|\) is the cardinality of the codeword symbol set, public as part of the protocol.

Unlike conventional block codes, the codebook is generated online rather than fixed in advance. This does not prevent the decoder from reconstructing the message-to-codeword mapping, since the encoding policy is public and \eqref{codeword generation} depends only on the decoder's current posterior, which both communicating parties can reproduce. We next describe how this posterior is updated.

\subsubsection{Key Generation and Optimal Transport}
This section instantiates the $\mathsf{KeyGen}$ algorithm of Definition~\ref{stegosystem}. We adapt from the optimal transport (OT) coupling mechanism of ArcMark~\cite{gilani2026arcmark} together with a transcript-dependent key generation procedure. To transmit \(u_t\), the encoder samples token \(x_t\) from the biased channel \(W_{X\mid U,K,S}(x_t \mid u_t,k_t,s_t)\), obtained by solving an OT problem subject to a distortion-free marginal constraint which we explain in detail in proceeding paragraphs.

Both encoder and decoder share a fixed secret \(\lambda\)-bit seed \(\kappa\), sampled uniformly once using a cryptographically secure random number generator (we use \(\lambda=128\) by default). Each LLM generation session is assigned a fresh public nonce \(\nu\sim\textnormal{Unif}(\{0,1\}^{\lambda})\), which is included in the shared session metadata. At token position \(t\), the key \(k_t\) and posterior-matching randomness \(R_t\) are generated by applying a keyed pseudorandom function to the nonce, the previous \(h\) token indices, and the current token position\footnote{In single-bit watermarking, side information is often derived by hashing a short window of previous tokens; see Appendix~A of \cite{tsur2026heavywater}. Incorporating the token position into the hash input eliminates repeated-window key collisions, but can make detection sensitive to cropping because the detector may not know the original token positions. This issue is less relevant in our multi-bit steganographic communication setting, where the encoder and decoder remain synchronized and decode the complete generated text. Another common remedy is to skip embedding whenever the same context window reappears.}
\begin{equation}
  \Lambda_t = F_{\kappa}\big(\nu \,\|\, x^{t-h:t-1} \,\|\, t\big), 
\end{equation}
where \(F_{\kappa}(z)\) is a keyed pseudorandom function. The context window is zero-padded for \(t\le h\) (we use \(h=3\) by default)~\cite{kirchenbauer2023watermark}. The nonce separates different generation sessions, while the position counter prevents repeated PRF inputs within a session. We split the resulting digest into three disjoint blocks
\(\Lambda_t^{(1)},\Lambda_t^{(2)},\Lambda_t^{(3)}\). The first two blocks
deterministically specify the channel index and the random permutation used
by the OT coupling. The third block is mapped by a fixed public conversion
function \(\mathsf{Rand}:\{0,1\}^{*}\to(0,1)\) to the synchronized
posterior-matching randomness $R_t$.
The complete side information is therefore $\left((k_t^{(1)},k_t^{(2)}), R_t\right)$
\begin{align}\label{eq:key-generation}
    & k_t^{(1)} = \mathrm{int}\big(\Lambda_t^{(1)}\big)\bmod r, \quad k_t^{(2)} = \Psi\big(\mathcal{X};\Lambda_t^{(2)}\big), \nonumber \\
    & R_t = \mathsf{Rand}\big(\Lambda_t^{(3)}\big),
\end{align}
where \(k_t=(k_t^{(1)},k_t^{(2)})\) is the key used to synthesize the biased channels and \(r=|\mathcal{K}^{(1)}|\) denotes the cardinality of the first key component so that
\(k_t^{(1)}\in\{0,\ldots,r-1\}\). \(R_t\in(0,1)\) is the dither used to generate codeword symbol. \(\Psi(\cdot\,;\,\Lambda_t^{(2)})\) denotes a deterministic permutation of the vocabulary parameterized by \(\Lambda_t^{(2)}\). Equation~\eqref{eq:key-generation} is precisely the $\mathsf{KeyGen}$ algorithm: given the shared seed $\kappa$, transcript history and public nonce $\nu$, it emits the per-token side information $(k_t,R_t)$ consumed by the encoder and decoder.

Following~\cite{gilani2026arcmark}, tokens and symbols are embedded on the unit circle: token \(x\) is assigned angle \(2\pi k^{(2)}_t(x)/V\) (permutation $k^{(2)}_t$ applied to the set $\mathcal{X}$), while the symbol \(u_t\) and \(k^{(1)}_t\) determine the target angle \(z_t = \big(2\pi u_t/p + 2\pi k^{(1)}_t/r + \phi\big)\bmod 2\pi\) for some fixed angle $\phi$. For realized $S_t=s_t$, the biased sampling channel is obtained by solving the transport plan $P_{X_t,Z_t} \in \mathbb{R}_+^{V\times r}$
\begin{align}\label{eq:ot}
    P^*_{X_t,Z_t} & = \arg\ \min_{P \in \mathbb{R}_+^{V\times r}} \sum_{i=0}^{V-1} \sum_{j=0}^{r-1} P_{i,j} \Phi_{i,j}\\
  &\quad \textnormal{s.t.} \quad \sum_{j=0}^{r-1}P_{i,j} = W_{X_t|S_t}(i|s_t),\quad \forall i \in [0,V-1],\nonumber\\
  &\quad \textnormal{s.t.} \quad \sum_{i=0}^{V-1}P_{i,j} = \frac{1}{r}, \quad \forall j \in [0,r-1],\nonumber
\end{align}
where the cost matrix $\Phi \in \mathbb{R}^{V\times r}$ is 
\begin{equation}
    \Phi_{i,j} = d\bigg(\frac{2\pi k^{(2)}_t(i)}{V},\, \big(\frac{2\pi u_t}{p} + \frac{2\pi j}{r} + \phi\big)\bmod 2\pi\bigg)
\end{equation}
In practice we solve (\ref{eq:ot}) using the Sinkhorn algorithm and recover the conditional distribution at the realized key via \(W_{X_t\mid U_t,K_t,S_t}(\cdot|u_t,k_t,s_t)=r\,P^*_{X_t,Z_t}(\cdot,z_t)\), with \(\Pr(Z_t=z_t)=1/r\). Details for the algorithm setup are provided in Appendix \ref{app:details}.

\subsubsection{Belief Computation and Encoding}
For each received token $x$, the decoder estimates the received symbol statistic \(\hat{C}_{t} = \big(2\pi k^{(2)}_t(x)/V  -2\pi k^{(1)}_t/r \big) \bmod 2\pi\) and evaluates a robust Laplace-contaminated likelihood,
\begin{align}\label{eq:mixture-likelihood}
    \ell_t(u_t(m)) &= f\!\left(d\!\left(\hat{C}_{t}, \tfrac{2\pi u_{t}(m)}{p}+\phi \right)\right) \nonumber\\
    &= \frac{1-\epsilon}{2b(1-e^{-\pi/b})} \exp\!\left(-\frac{d\!\left(\hat{C}_{t},\, \frac{2\pi u_t(m)}{p} + \phi\right)}{b}\right) + \frac{\epsilon}{2\pi},
\end{align}
with \(b = \pi/(p\sqrt{2})\) and \(\epsilon \in [0,1)\)~\cite{chen2016general}. The Laplace component models the received symbol as concentrated around the transmitted codeword symbol, while the uniform component accounts for outlier tokens whose decoded symbol statistic carries little information. The decoder then updates its posterior according to
\begin{equation}\label{belief update}
    \pi_t(m) = \frac{\pi_{t-1}(m)\,\ell_t(u_t(m))}{\sum_j \pi_{t-1}(j)\,\ell_t(u_t(j))}.
\end{equation}
We note that for each message $m$, the mapping to codeword symbol $u_t(m)$ at position $t$ is deterministic and available to both parties through (\ref{codeword generation}). 

\subsubsection{Decoding}
Because both communicating parties observe the same generated transcript, they maintain identical posterior beliefs throughout communication. Decoding therefore mirrors the encoding procedure. The decoder can declare a tentative message whenever there exists a message whose belief crosses threshold $\gamma$, $\exists m',\pi_t(m')>\gamma$. In Algorithm~\ref{alg:bam} we augment the decoding with a two-phase threshold decoding.\\

A visual illustration of the posterior matching belief is shown in Appendix \ref{app:algorithms}. We also present the encoder and decoder as a single mirrored procedure in Appendix \ref{app:algorithms}. 

\subsection{The BAM Protocol}\label{sec:bam}
Posterior matching progressively concentrates the shared belief, but the leading candidate can still be incorrect. Once one candidate dominates, BAM treats verification as a binary ACK/NACK test and repeatedly transmits antipodal confirmation symbols. This design is inspired by the two-phase Yamamoto--Itoh strategy ~\cite{yamamoto1979asymptotic,burnashev1976data}. We now describe the complete BAM protocol, using Algorithm~\ref{alg:pm} as its communication phase. The BAM protocol is parameterized by three thresholds,
\(\gamma\), \(\gamma_{\textnormal{ACK}}\), and \(\gamma_{\textnormal{NACK}}\), satisfying
\(0.5\leq\gamma, \gamma_{\textnormal{ACK}}, \gamma_{\textnormal{NACK}}<1\). These parameters are part of the publicly known policy and therefore require no synchronization.

\begin{enumerate}[leftmargin=1.3em,itemsep=2pt,topsep=2pt]

\item During the communication phase, the encoder generates codeword symbols and tracks the decoder's posterior according to \eqref{codeword generation} and \eqref{belief update} using Algorithm~\ref{alg:pm}.

\item Whenever the decoder's posterior satisfies
\(\pi_t(m')>\gamma\) for some \(m'\in\mathcal M\), the protocol enters the confirmation phase and encoder transmits one of two confirmation symbols,
\(u_{\mathrm{ACK}}=0\) if $m'$ is the true message or
\(u_{\mathrm{NACK}}=1\) otherwise
(corresponding to \(\theta_{\mathrm{ACK}}=2\pi u_{\mathrm{ACK}}/p=0\) and \(\theta_{\mathrm{NACK}}=2\pi u_{\mathrm{NACK}}/p=\pi\) with \(p=2\)).

\item During confirmation, the encoder tracks the decoder's posterior over
\(\{u_{\mathrm{ACK}},u_{\mathrm{NACK}}\}\) by applying belief update in (\ref{eq:mixture-likelihood}), (\ref{belief update}) with \(p=2\) and likelihood parameter $\epsilon_{\mathrm{ACK}} \in [0,1)$. The confirmation phase terminates once either
\(\pi_t(u_{\mathrm{ACK}})>\gamma_{\textnormal{ACK}}\), in which case the tentative message $m'$ is accepted and the transmission terminates, or
\(\pi_t(u_{\mathrm{NACK}})>\gamma_{\textnormal{NACK}}\), in which case the tentative message is rejected, the decoder reduces the posterior of the candidate message by a factor
\((1-\gamma_{\textnormal{NACK}})/\gamma_{\textnormal{NACK}}\), renormalizes the posterior, and protocol re-enters communication phase.

\item \textbf{False Reject.}
If
\(\pi_t(u_{\mathrm{NACK}})>\gamma_{\textnormal{NACK}}\)
even though \(u_{\mathrm{ACK}}\) was transmitted, the protocol re-enters the communication phase and continues to transmit the message via Algorithm~\ref{alg:pm}.

\item \textbf{False Accept.}
If
\(\pi_t(u_{\mathrm{ACK}})>\gamma_{\textnormal{ACK}}\)
when \(u_{\mathrm{NACK}}\) was transmitted, the protocol declares a decoding error event.

\item \textbf{Timeout.}
If no message is accepted by token \(T^*\), the decoder outputs
\(\hat{m}=\arg\max_{m'\in\mathcal M}\pi_{T^*}(m')\).

\end{enumerate}

The algorithm is therefore specified by the parameter tuple
\begin{equation}\label{BAMparameters}
    \mathrm{BAM}(\gamma,\gamma_{\textnormal{ACK}},\gamma_{\textnormal{NACK}},(\epsilon,\epsilon_{\textnormal{ACK}}),T^*),
\end{equation}
where
\((\epsilon,\epsilon_{\textnormal{ACK}})\)
are the likelihood parameters in (\ref{eq:mixture-likelihood}) used during the communication and confirmation phases, respectively. The complete procedure is summarized in Algorithm~\ref{alg:bam} in Appendix~\ref{app:algorithms}.

The two branches of Algorithm~\ref{alg:pm} and \ref{alg:bam} instantiate the remaining components of Definition~\ref{stegosystem}. The encoder branch is $\mathsf{Enc}_{\mathsf{Model},K}$: given the message $m$, prompt $\rho$, and the side information $(k_t,R_t)$ from $\mathsf{KeyGen}$, it forms the codeword symbol $u_t$ by \eqref{codeword generation} and samples the token $x_t$ from the biased channel \eqref{eq:ot}, appending it to the stegotext $x^\tau$. The decoder branch is $\mathsf{Dec}_K$: observing only $x^\tau$ and the shared side information, it computes $u_t(m')$ for every candidate $m'$, scores the likelihood \eqref{eq:mixture-likelihood}, and updates the belief \eqref{belief update} to produce the estimate $\hat m$. A complete, per-party account of the information available in the system is given in Appendix~\ref{app:party-info}.

\begin{remark}[Confirmation phase as binary hypothesis testing]
    The confirmation phase is equivalent to a variable length repetition code over anti-podal symbols. The posterior over
    \(\{u_{\mathrm{ACK}},u_{\mathrm{NACK}}\}\)
    is updated using the contaminated-Laplacian likelihood in (\ref{eq:mixture-likelihood}) with \(p=2\) and a separate contamination parameter
    \(\epsilon_{\mathrm{ACK}}\). Since binary antipodal signaling experiences lower effective channel noise than the communication phase, the protocol does not require $\epsilon=\epsilon_{\mathrm{ACK}}$. Unlike Algorithm \ref{alg:pm}, in the confirmation phase, the posterior is used only in decoding, which is a binary hypothesis testing problem. 
\end{remark} 
\begin{remark}[Weights of type I and II error]          
    A false reject incurs only an additional retransmission, whereas a false accept results in an unrecoverable decoding error. Consequently, we choose
\(\gamma_{\textnormal{ACK}}>\gamma_{\textnormal{NACK}}\),
biasing the protocol toward retransmission rather than erroneous acceptance. 
\end{remark}

\section{Security Analysis}\label{sec:security}
We first establish the formal security guarantee for BAM through a computational indistinguishability proof against a polynomial-time adaptive API adversary. We then empirically validate marginal preservation under the implemented key generation.

\subsection{Cryptographic Security}\label{sec:crypto-security}

The formal guarantee targets the security definition of Definition~\ref{crypto}. The adversary samples one fixed secret seed \(\kappa\) and keeps it hidden, while every API query receives a fresh public nonce. We assume that the adversary makes at most \(Q(\lambda)\) queries and observes at most \(L(\lambda)\) generated tokens. Note that in our algorithm, the decoder commits to a message decoding at an internal codeword length $n\leq \tau$, the token generation continues under the same marginal-preserving sampler until natural generation stop $\tau$. Therefore, as will be shown, the length of generation does not enable adversary to distinguish covertext from stegotext.

\begin{theorem}\label{crypto proof}
Let \(F_{\kappa}\) be the PRF used to generate per-token side information. For every polynomial-time distinguishing adversary \(\mathcal A\) making at most \(Q(\lambda)\) queries and observing at most \(L(\lambda)\) tokens, there exists a PRF distinguisher \(\mathcal B\) such that
\begin{equation}\label{reduction}
    \operatorname{Adv}_{\mathcal A}^{\mathrm{BAM}}(\lambda)
    \le
    \operatorname{Adv}_{\mathcal B}^{\mathrm{PRF}}(\lambda)
    +
    \frac{Q(\lambda)(Q(\lambda)-1)}{2^{\lambda+1}}
    +
    L(\lambda)\eta(\lambda),
\end{equation}
where \(\operatorname{Adv}_{\mathcal B}^{\mathrm{PRF}}\) denotes the distinguishing advantage against the underlying PRF and \(\eta(\lambda)\) is the per-token deviation from the ideal OT marginal constraint under a truly random function, complete definition provided in Appendix~\ref{app:crypto-proof}.
\end{theorem}

\begin{proof}
Sketch.
The proof follows a standard sequence-of-games argument. We first replace the PRF used to derive BAM side information with a truly random function. Any adversary capable of distinguishing these two games immediately yields a distinguisher against the underlying PRF, establishing the first term in~\eqref{reduction}. On the other hand, under an idealized key schedule, the OT marginal constraint guarantees that every generated token has the same distribution as under the unbiased language model. We then couple the BAM and unbiased generation processes token by token. A birthday bound yields the second term in~\eqref{reduction}, while deviations from the ideal OT marginal condition contribute to the third term \(L(\lambda)\eta(\lambda)\). Combining the reduction with the coupling argument completes the proof. The full proof is provided in Appendix~\ref{app:crypto-proof}.
\end{proof}

We stress that although BAM employs a variable internal decoding time $n$, this stopping time does not determine the observable generation length. Each conversational turn terminates only when EOS is sampled from the LLM. Moreover, since each query results in a finite-length token generation, both $Q(\lambda)$ and $L(\lambda)$ are polynomial in $\lambda$. Consequently, the distinguishing advantage in~\eqref{reduction} is negligible whenever the underlying PRF is secure and $\eta(\lambda)$ is negligible. For the ideal OT construction enforced in (\ref{eq:ot}) and analyzed in Theorem~\ref{crypto proof}, the base distribution is imposed as an exact marginal constraint, so $\eta(\lambda)=0$. The empirical study of Section~\ref{sec:empirical-marginal} instead evaluates the realized finite-precision implementation, accounting for both the PRF-induced key distribution and numerical deviation of the OT solver.

\subsection{Empirical Validation of Marginal Preservation}\label{sec:empirical-marginal}

For every token position, the OT construction makes the biased channel
($W_{X\mid U,K,S}(\cdot\mid u,k,s)$) marginalize over key to ($W_{X\mid S}(\cdot\mid s)$). This identity holds by construction for coupling-based schemes~\cite{de2022perfectly}. In practice, however, two sources of deviation may arise: the realized key distribution is induced by a PRF rather than sampled directly from a uniform key, and the OT coupling is computed numerically to finite precision. We therefore estimate the seed-marginalized per-token divergence\
\begin{equation}
    \mathbb D\big(W_{X|S}(x|s)\big|\big|\sum_{k\in \mathcal K}P_K(k)W_{X|U,K,S}(x|u,k,s)\big),
\end{equation}
where $P_K$ is the per-token key distribution induced by the realized PRF schedule under a uniform seed. We note that by Definition~\ref{cachin}, we use a per-token KL constraint for every input symbol $u$ and token history, which is stronger constraint than \cite[Definition 2]{cachin2004information}.

Fig~\ref{fig:KL} in Appendix estimates Definition~\ref{cachin} by Monte Carlo over an increasing
number of seeds. At every token length, the estimate descends toward the sampling floor
as the seed count grows, while the accumulated divergence grows roughly linearly in token length. $10^5$ sampled seeds result in an empirical per-token KL smaller than $10^{-5}$. Therefore, the experiment found no local deviation resolvable at the reported Monte Carlo precision. 
The realized schedule is thus empirically \(\sigma\)-secure on average with small \(\sigma\) at the tested seed counts (Definition~\ref{cachin}). Figure~\ref{fig:KL} should be viewed as an implementation diagnostic capturing both PRF-induced key non-uniformity and finite-precision OT solver.

\section{Covert Agentic Conversation}\label{sec:convo}
Section~\ref{sec:method} describes the BAM algorithm, which can be applied to any general LLM generated cover. However, conversational covertext in particular creates additional challenges not present in long, continuous generative text. In this section we explain several design considerations of the BAM algorithm specifically for conversational steganography. 

A defining feature of covert agentic communication, compared with conventional covert communication over static covertexts~\cite{bloch2016covert,chen2023covert}, is that the deployer controls the \emph{interaction} but not the \emph{length} of the covertext. The conversational setting, agent roles, and turn structure are chosen in advance, yet under the distortion-free constraint the language model cannot be forced to terminate at a prescribed token or continue once an EOS token is sampled. Consequently, we may resolve this mismatch by adapting BAM's decoding decision to the realized generation rather than to a fixed belief threshold. This flexibility also makes the covertext itself a design variable: conversational settings with longer or higher-entropy turns naturally provide greater embedding opportunities (how to design suitable prompts for conversational steganography is outside the scope of this study). We first describe the agentic conversation setup considered in this work.

\paragraph{Agentic conversation setup.}

Both agents are first provided with a \emph{shared background description} specifying the conversational setting, assigning each agent a role, and an opening. Each agent additionally receives a private \emph{instruction} governing the content and style of its responses. For automation, we keep these instructions fixed throughout a conversation, although they may be customized on a turn-by-turn basis according to user inputs. The effective context for each generation therefore consists of the shared background, the agent's private instructions, and the public conversation history. Example conversation templates are provided in Appendix \ref{app:covert-results}.

\paragraph{Adaptive stopping.}

To improve token efficiency in conversational steganography, we consider the following modification to BAM, which adapts decoding to the realized conversational turn rather than enforcing a fixed decoding threshold.

\begin{enumerate}
    \item \textbf{Case I: $n\leq\tau$.} In this case, the codeword is completed at a token budget $n$ shorter than natural generation length $\tau$. The posterior belief exceeds the default decoding communication and confirmation thresholds \((\gamma,\gamma_{\textnormal{ACK}})\) before EOS is sampled, the encoder continues transmitting the message until EOS, while the decoder makes use of the additional tokens and chooses the highest belief message until the turn naturally terminates at EOS, thus creating better channel utilization. 
    \item \textbf{Case II: $n>\tau$.} In this case, the codeword cannot be completed before the natural generation $\tau$ of a conversation turn. If EOS is sampled before the confirmation phase completes, generation terminates normally instead of being artificially extended. Specifically, since the decoder's belief in \(u_{\mathrm{ACK}}\) remains below \(\gamma_{\textnormal{ACK}}\) (including cases when confirmation phase is not reached), the opposing agent produces a genuine, unbiased response that carries no payload, allowing the overt conversation to proceed naturally. The transmitting agent then resumes embedding the remaining payload during its next turn.
\end{enumerate}

\paragraph{Interactive composition and security}
By allowing the stopping time to follow the natural generation process, rather than a predetermined token budget, the modified algorithm can simultaneously achieve a higher throughput and better token utilization while preserving the distortion-freeness from point-to-point communication. 
Note that for multi-turn steganography under our construction, the security guarantee in Theorem~\ref{crypto proof} is local to each communication turn and is conditioned on the public conversation history. Alternating the point-to-point BAM transmissions therefore compose: if each communication turn is computationally indistinguishable from clean generation conditioned on every preceding public transcript, then a standard hybrid argument over any polynomial number of turns shows that the complete two-way conversation is computationally indistinguishable from the corresponding clean conversation.

\section{Evaluation}\label{sec:experiment}

\subsection{Overview}

We adopt a layered evaluation framework. The first layer evaluates BAM as a token-level steganographic embedding primitive on standardized continuous-text covers. These experiments measure its rate--reliability tradeoff, text quality and computational cost against existing multi-bit watermarking baselines. The second layer evaluates the composition of this same primitive into an end-to-end covert agentic communication protocol. To the best of our knowledge, such evaluation benchmarks and corresponding baselines do not currently exist. Therefore, rather than repeating primitive-level measurements, we propose an evaluation framework for covert conversations that focuses on properties that arise only through multi-turn composition, including payload throughput, dialogue-level reliability, token utilization and conversational quality.

For continuous-text experiments, we use three base models: Llama-3.1-8B, Qwen-3.5-9B-Base, and Mistral-7B-v0.3. For covert agentic communication, we use Llama-3.1-8B-Instruct, Phi-4-14B-Instruct, Qwen3-A3B-30B-Instruct-2507. All models are accessed at inference time using temperature \(1.0\) and top-\(50\) sampling. Experiments run with batch size \(1\) on a single NVIDIA A100-SXM4 GPU with \(80\,\mathrm{GB}\) memory, \(4\) CPU cores, and \(64\,\mathrm{GB}\) system memory on a SLURM cluster.

Throughout, \(F_U^{-1}\) is the inverse CDF of the uniform distribution, and posterior updates use the mixture-Laplace likelihood in (\ref{eq:mixture-likelihood}). Codeword symbols are drawn from an alphabet of size \(p=|\mathcal U|=4\) with \(\phi=0\), and the channel-key cardinality is \(r=|\mathcal K^{(1)}|=4\). We use a \(128\)-bit security parameter \(\lambda=128\), context-window length \(h=3\), and $F_{\kappa}(z)=\operatorname{HMAC\text{-}SHA256}(\kappa,z)$.

\subsection{Evaluation Metrics}

\subsubsection{Primitive-Level Metrics}

The continuous-text experiments use the C4 Real News dataset and evaluate BAM against existing inference-time multi-bit watermarking schemes.

\begin{itemize}
    \item \textbf{Reliability and Rate.}
    We evaluate the empirical counterpart of Definition~\ref{def:reliability} using the average message-decoding error probability over messages and prompts sampled uniformly across \(1000\) trials. Message error rate in Definition~\ref{def:reliability} is stricter than the bit-error rate reported by several baselines~\cite{feng2025bimark,cui2026mc}: a message is counted as incorrect whenever at least one payload bit is decoded incorrectly. This is the more operationally relevant metric for communication setting when no outer error-correction code is used. The effective embedding rate is defined in Definition~\ref{rate}.
    \item \textbf{Text Quality.}
    We evaluate fluency using perplexity and diversity using distinct-\(n\). Perplexity is scored against full-softmax distribution and distinct-$n$ measures the percentage of unique $n$-grams. Because BAM is variable-length, every stego generation is compared with a length-matched no-embedding generation from the same base model. A smaller perplexity difference indicates better preservation of fluency, while a larger distinct-\(n\) value indicates less repetition.
    \item \textbf{Generation Time.}
    We report wall-clock generation time in seconds per token. Measurements are CUDA-synchronized and collected after untimed warm-up generations. Posterior-based decoding scales exponentially with the payload size; for BAM, the synchronized posterior affects both encoding and decoding. A detailed runtime breakdown and scaling analysis are provided in Appendix~\ref{app:extra}.
\end{itemize}

\subsubsection{Conversation-Level Metrics}

As described in Section~\ref{sec:convo}, for the conversational setting we simulate \(100\) dialogues of \(10\) rounds each, with each dialogue carrying up to \(10\) hidden payloads. The covert tasks are stateful in nature, therefore each round can carry only one payload. We focus on end-to-end properties introduced by variable turn lengths, alternating speakers, and multi-turn task dependence. Formal definitions are provided in Appendix \ref{app:covert-setup}. 

\begin{itemize}
    \item \textbf{Bits/token}: The average number of successfully decoded payload bits per generated token. This measures the effective communication rate under the lower average entropy and variable turn lengths of conversational text.

    \item \textbf{Avg payload/dialogue}: The average number of successfully transmitted payloads per conversation. This captures how much of the hidden task can be completed within the available dialogue.

    \item \textbf{Message and dialogue error rate}: Message error rate measures individual payload failures. Dialogue error rate is the fraction of ten-round conversations containing at least one incorrect payload, measuring the reliability required to complete an entire multi-message conversational session without communication error.

    \item \textbf{Efficiency}: The fraction of payload-carrying tokens. This measures utilization of the naturally available conversational cover.

    \item \textbf{Conversational Quality}: We use a blinded LLM judge to compare BAM and clean conversations with respect to naturalness, response relevance, multi-turn coherence, role adherence, repetition, and overt-task consistency. Transcripts are presented in randomized order, and we report the proportions of BAM wins, ties, and losses against length- and prompt-matched clean conversations.
\end{itemize}

\begin{figure*}[t]
    \centering
    \includegraphics[width=\linewidth]{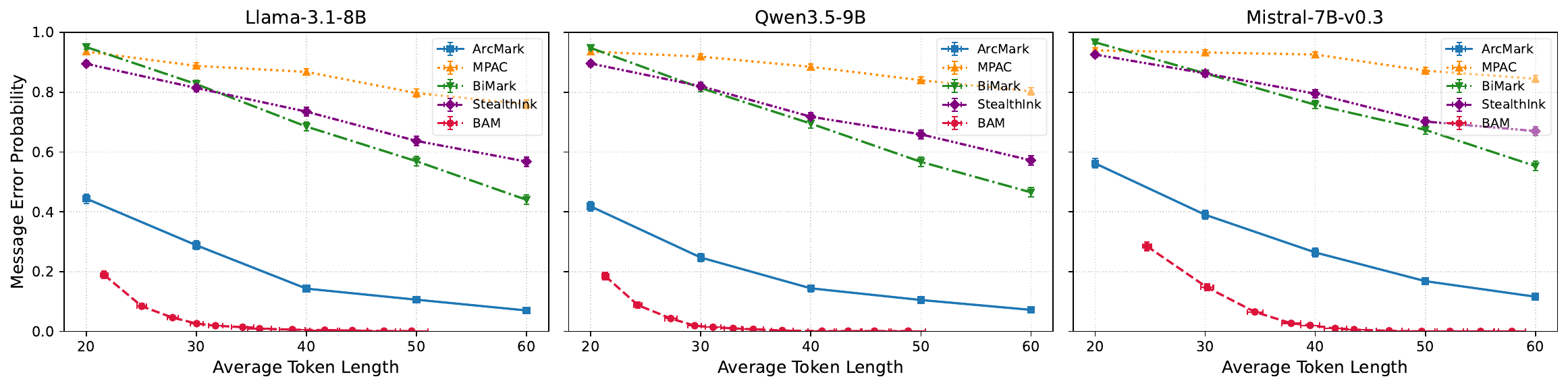}
    \caption{BAM versus baseline schemes on the C4 News dataset with an 8-bit payload, across three base models (Llama-3.1-8B, Qwen-3.5-9B, Mistral-7B-v0.3). Each
    point trades message-decoding error against average token length; BAM (with SEM on length)
    attains significantly lower error at shorter average lengths than other baselines.}
    \label{fig:bam_vs_fl}
\end{figure*}

\begin{table*}[t]
\centering
\small
\setlength{\tabcolsep}{4.5pt}
\renewcommand{\arraystretch}{1.15}
\caption{
Comparison of BAM scheme vs Baselines on C4 News Dataset.
Reliability is reported as average generation length and message error rate
(lower error is better). Values are mean\,\(\pm\)\,SE over 1000 trials.
Models: Llama: Llama-3.1-8B; Qwen: Qwen-3.5-9B-Base; Mistral: Mistral-7B-v0.3}
\label{tab:mega_comparison}
\begin{tabular}{llccccccc}
\toprule
\multirow{2}{*}{\textbf{Model}} & \multirow{2}{*}{\textbf{Scheme}}
& \multicolumn{2}{c}{\textbf{Reliability}}
& \multicolumn{4}{c}{\textbf{Text Quality}}
& \multicolumn{1}{c}{\textbf{Time}} \\
\cmidrule(lr){3-4}\cmidrule(lr){5-8}\cmidrule(lr){9-9}
& & \textbf{Avg Len} & \textbf{Msg Err} \(\downarrow\)
  & \textbf{PPL} \(\downarrow\) & \textbf{Dist2} & \textbf{Dist3} & \textbf{Dist4}
  & \textbf{Sec/Token} \(\downarrow\)\\
\midrule
\multirow{7}{*}{\makecell[l]{\textbf{Llama}}}
& No Embedding & 50.0 & -- & 7.66\,$\pm$\,0.11 & 0.969 & 0.986 & 0.992 & 0.0214 \\
& MPAC         & 50.0 & 0.797\,$\pm$\,0.013 & 8.54\,$\pm$\,0.11  & 0.963 & 0.982 & 0.989 & 0.0223 \\
& BiMark       & 50.0 & 0.569\,$\pm$\,0.016 & 7.68\,$\pm$\,0.11  & 0.959 & 0.977 & 0.984 & 0.0352 \\
& StealthInk   & 50.0 & 0.637\,$\pm$\,0.015 & 7.66\,$\pm$\,0.11  & 0.969 & 0.987 & 0.992 & 0.0231 \\
& ArcMark      & 50.0 & 0.106\,$\pm$\,0.010 & 7.52\,$\pm$\,0.11  & 0.972 & 0.988 & 0.994 & 0.0291 \\
\cmidrule(lr){2-9}
& No Embedding    & 47.1\,$\pm$\,1.5 & -- & 7.71\,$\pm$\,0.12 & 0.971 & 0.987 & 0.992 & 0.0224 \\
& BAM(L=$2^{15}$) & 47.1\,$\pm$\,1.5 & \textbf{0.001\,$\pm$\,0.001} & 7.89\,$\pm$\,0.12 & 0.970 & 0.984 & 0.989 & 0.0378 \\
\midrule
\multirow{7}{*}{\makecell[l]{\textbf{Qwen}}}
& No Embedding & 50.0 & -- & 7.83\,$\pm$\,0.10 & 0.966 & 0.989 & 0.994 & 0.0420 \\
& MPAC         & 50.0 & 0.840\,$\pm$\,0.012 & 9.03\,$\pm$\,0.12  & 0.968 & 0.989 & 0.995 & 0.0420 \\
& BiMark       & 50.0 & 0.567\,$\pm$\,0.016 & 8.23\,$\pm$\,0.12  & 0.957 & 0.980 & 0.987 & 0.0558 \\
& StealthInk   & 50.0 & 0.659\,$\pm$\,0.015 & 8.04\,$\pm$\,0.11  & 0.965 & 0.989 & 0.995 & 0.0456 \\
& ArcMark      & 50.0 & 0.105\,$\pm$\,0.010 & 7.76\,$\pm$\,0.11  & 0.963 & 0.987 & 0.993 & 0.0510 \\
\cmidrule(lr){2-9}
& No Embedding    & 45.8\,$\pm$\,1.3 & -- & 8.21\,$\pm$\,0.13 & 0.970 & 0.989 & 0.995 & 0.0436 \\
& BAM(L=$2^{15}$) & 45.8\,$\pm$\,1.3 & \textbf{0.001\,$\pm$\,0.001} & 8.24\,$\pm$\,0.12 & 0.969 & 0.988 & 0.993 & 0.0541 \\
\midrule
\multirow{7}{*}{\makecell[l]{\textbf{Mistral}}}
& No Embedding & 50.0 & -- & 6.25\,$\pm$\,0.09 & 0.962 & 0.985 & 0.992 & 0.0212 \\
& MPAC         & 50.0 & 0.872\,$\pm$\,0.012 & 6.73\,$\pm$\,0.09  & 0.959 & 0.981 & 0.989 & 0.0216 \\
& BiMark       & 50.0 & 0.674\,$\pm$\,0.015 & 6.26\,$\pm$\,0.08  & 0.958 & 0.980 & 0.988 & 0.0342 \\
& StealthInk   & 50.0 & 0.702\,$\pm$\,0.014 & 6.16\,$\pm$\,0.08  & 0.963 & 0.986 & 0.994 & 0.0217 \\
& ArcMark      & 50.0 & 0.168\,$\pm$\,0.012 & 6.13\,$\pm$\,0.09  & 0.959 & 0.982 & 0.991 & 0.0278 \\
\cmidrule(lr){2-9}
& No Embedding    & 49.7\,$\pm$\,1.2 & -- & 6.22\,$\pm$\,0.10 & 0.960 & 0.981 & 0.989 & 0.0222 \\
& BAM(L=$2^{11}$) & 49.7\,$\pm$\,1.2 & \textbf{0.000\,$\pm$\,0.000} & 6.51\,$\pm$\,0.09 & 0.963 & 0.984 & 0.991 & 0.0326 \\
\bottomrule
\end{tabular}
\end{table*}

\subsection{Performance on Continuous Text} \label{contin}
\paragraph{Setup} We first demonstrate BAM's steganography performance on the C4 Real News dataset. We compare against benchmark schemes under multi-bit watermarking  ArcMark~\cite{gilani2026arcmark}, MPAC~\cite{yoo2023advancing},
BiMark~\cite{feng2025bimark} and StealthInk~\cite{jiang2025stealthink}, the specific experimental setup details are provided in Appendix~\ref{app:details}\footnote{We are aware that there are other distortion-free multi-bit watermark works such as \cite{cui2026mc,jiang2026mirrormark}. However, they do not currently provide a public implementation, so we omit them from benchmarking. Among the baselines, MPAC is not distortion-free, but we include it as a reference since it is a seminal work.}. For algorithm setting, we use
BAM\((0.5,\,1-L^{-1},\,0.75,\,(0.4,0.4),\,1000)\) for (\ref{BAMparameters}) while sweeping the confirmation threshold 
\(L \in \{2,2^2,2^3,2^4,2^5,2^6,2^7,2^9,2^{11},2^{13},2^{15}, 2^{17}\}\). For each setting, we conduct \(1000\) trials. We embed an $8$-bit payload across all settings. Because BAM's length is variable, each
watermarked scheme is compared against its \emph{own} length-matched no-embedding baseline.

It is important to note that communication systems commonly employ sub-packetization to balance computational complexity and error performance. For example, a $64$-bit message may be divided into eight $8$-bit packets, substantially reducing the posterior-update complexity. We therefore focus on $8$-bit payloads, while additional performance and runtime results for longer payloads are provided in Appendix~\ref{app:extra}.

\paragraph{Evaluation} 
The full reliability and average token length tradeoff is reported in Fig.~\ref{fig:bam_vs_fl}, and we expand the evaluation on other performance measures in Table~\ref{tab:mega_comparison}.
\begin{itemize}
    \item \textbf{Reliability}: BAM achieves significantly smaller error at shorter average length and the performance gain is roughly two orders of magnitude better than the best baseline. Across all three models BAM achieves a substantially better reliability-length
    tradeoff than the benchmark baselines. 
      Concretely, BAM attains a message error rate of \(\leq 0.1\%\) over \(1000\) runs with an average token below \(50\) across all three LLM models, whereas
the strongest baselines of comparable
length incur \(10\)--\(17\%\) message error. The strong reliability guarantee is the cornerstone of ensuring covert communication. A detailed ablation for explaining the sources of performance gain is provided in Appendix~\ref{app:extra}.
    \item \textbf{Text Quality}: In these trials, the stegotext generated by our method retains similar linguistic quality of texts generated from a clean model. The BAM outputs show only a modest perplexity
    difference across three models, compared to the base perplexity. Such differences are on par with other distortion-free algorithms such as \cite{jiang2025stealthink,gilani2026arcmark,feng2025bimark}, while exhibiting better quality than distortionary \cite{yoo2023advancing}.
    \item \textbf{Time}: The additional runtime of BAM comes mainly from channel synthesis (OT) and posterior update. As shown in Table~\ref{tab:mega_comparison}, BAM runtime is the weakest part of the algorithm and exhibits a $30\%,6\%,17\%$ increase compared to ArcMark over Llama, Qwen and Mistral, respectively. This is a result of posterior matching, which must be computed at both encoder and decoder. A detailed breakdown of runtime, as well as scalability concerns, are addressed in Appendix~\ref{app:extra}.
\end{itemize}

\subsection{Performance on Covert Agentic Communication}\label{non-contin} 
\paragraph{Setup} We test the covert communication
system in a two-way conversational setting. We use three specific turn-based covert tasks, which differ in the size of the hidden payload transmitted per message: a tic-tac-toe move (avg.\ 2.3 coded bits), a chess move (avg.\ 5.0 coded bits), and a move in $19\times19$ Go (\texttt{go19}, avg.\ 8.4 coded bits). We use uniform-prior source coding over the legal moves available in the current position \footnote{The performance of source coding can be improved via Huffman encoding when taking into account the distribution of legal moves. However, optimizing source coding or prompting is beyond the scope of this paper.}. We use two conversational covers: casual text-message chats and forum exchanges. In general, casual chats provide shorter conversation length, resulting in a lower overall embedding capacity. The shared background descriptions are generated by prompting GPT-5.5 to produce a pool of $50$ distinct scenarios for each setting \cite{hagendorff2024deception}. Each dialogue samples one background from its setting's pool, so the covert channel is evaluated across a distribution of topics, while the side instructions are fixed for each setting. For each combination of covert task and
overt conversation cover, we simulate $100$ conversations of $10$ rounds each, using the covert protocol
with setting BAM\((0.5, 0.75, 0.75, (0.7,0.3), 1000)\).
The default acknowledgment threshold $\gamma_{\mathrm{ACK}}$ is chosen to be lower than in the primitive test because we used the adaptive threshold rule designed for conversational cover discussed in Section ~\ref{sec:convo}, which improves channel efficiency at the cost of a marginal increase in decoding error.

Beyond reliability, the covert conversation must not induce quality degradation over unbiased conversations. Although we already demonstrated the quality in preceding sections, conversation setup may introduce multi-turn structure unaccounted for by primitive-level metrics. 
We generate $100$ stegotext conversations with a clean conversation generated from the same topic, opener, and sampling seed, and ask an independent LLM
judge (Gemini-3.1-Pro-Preview, Accessed Aug 10-24) which of the two is higher quality. This is intended to show that stegotext does not degrade the relative quality of multi-turn generation compared to unbiased conversation. The judge sees the pair in blinded and randomized order and  returns a win/tie/loss preference for the embedded transcript against its seed-matched clean counterpart. Additional experimental details provided in Appendix \ref{app:covert-results}.

\begin{table*}[t]
\centering
\caption{Turn-based covert channel on the casual chat covertext, across three open-weight instruct models and three embedding tasks. Values are mean $\pm$ SE over 100 dialogues per cell.}
\label{tab:turnbased_chat}
\setlength{\tabcolsep}{8pt}
\renewcommand{\arraystretch}{1.15}
\begin{tabular}{l l ccc}
\toprule
\multicolumn{2}{l}{\textbf{casual chat covertext}}
 & \multicolumn{3}{c}{\textbf{Embedding task}} \\
\cmidrule(lr){1-2}\cmidrule(lr){3-5}
\textbf{Model} & \textbf{Metric}
 & \makecell{\textbf{tictactoe}\\ \footnotesize avg. 2.3 bits}
 & \makecell{\textbf{chess}\\ \footnotesize avg. 5.0 bits}
 & \makecell{\textbf{go19}\\ \footnotesize avg. 8.4 bits} \\
\midrule
\multirow{6}{*}{\makecell[l]{\textbf{Llama-3.1-}\\ \textbf{8B-Instruct}\\[2pt] \footnotesize avg.\ entropy \\ \footnotesize 1.878 bits/tok,\\ \footnotesize len 33.7 tok}}
 & Bits/token           & 0.0607 $\pm$ 0.0011 & 0.0903 $\pm$ 0.0023 & 0.1030 $\pm$ 0.0029 \\
 & Avg payload/dialogue & 7.27 $\pm$ 0.19     & 5.20 $\pm$ 0.17     & 2.64 $\pm$ 0.09 \\
 & Dialogue error rate  & 0.050 $\pm$ 0.022   & 0.020 $\pm$ 0.014   & 0.060 $\pm$ 0.024 \\
 & Message error rate   & 0.0069 $\pm$ 0.0029 & 0.0038 $\pm$ 0.0032 & 0.0227 $\pm$ 0.0106 \\
 & Efficiency (\%)      & 88.64 $\pm$ 0.99    & 79.33 $\pm$ 0.89    & 65.25 $\pm$ 0.62 \\
 & Quality (w/t/l)      & \multicolumn{3}{c}{58 / 0 / 42} \\
\midrule
\multirow{6}{*}{\makecell[l]{\textbf{Phi-4-}\\ \textbf{14B-Instruct}\\[2pt] \footnotesize avg.\ entropy \\ \footnotesize 2.049 bits/tok,\\ \footnotesize len 62.3 tok}}
 & Bits/token           & 0.0394 $\pm$ 0.0008 & 0.0666 $\pm$ 0.0010 & 0.1028 $\pm$ 0.0022 \\
 & Avg payload/dialogue & 9.33 $\pm$ 0.12     & 8.35 $\pm$ 0.17     & 5.54 $\pm$ 0.21 \\
 & Dialogue error rate  & 0.020 $\pm$ 0.014   & 0.020 $\pm$ 0.014   & 0.070 $\pm$ 0.026 \\
 & Message error rate   & 0.0021 $\pm$ 0.0016 & 0.0024 $\pm$ 0.0017 & 0.0126 $\pm$ 0.0058 \\
 & Efficiency (\%)      & 98.08 $\pm$ 0.50    & 95.25 $\pm$ 0.65    & 84.89 $\pm$ 0.97 \\
 & Quality (w/t/l)      & \multicolumn{3}{c}{50 / 1 / 49} \\
\midrule
\multirow{6}{*}{\makecell[l]{\textbf{Qwen3-A3B-}\\ \textbf{30B-Instruct}\\[2pt] \footnotesize avg.\ entropy \\ \footnotesize 1.171 bits/tok,\\ \footnotesize len 37 tok}}
 & Bits/token           & 0.0434 $\pm$ 0.0010 & 0.0478 $\pm$ 0.0017 & 0.0470 $\pm$ 0.0030 \\
 & Avg payload/dialogue & 5.26 $\pm$ 0.21     & 3.25 $\pm$ 0.14     & 1.60 $\pm$ 0.07 \\
 & Dialogue error rate  & 0.190 $\pm$ 0.039   & 0.200 $\pm$ 0.040   & 0.220 $\pm$ 0.046 \\
 & Message error rate   & 0.0380 $\pm$ 0.0091 & 0.0708 $\pm$ 0.0197 & 0.0938 $\pm$ 0.0406 \\
 & Efficiency (\%)      & 79.53 $\pm$ 1.07    & 70.94 $\pm$ 0.81    & 62.56 $\pm$ 0.34 \\
 & Quality (w/t/l)      & \multicolumn{3}{c}{46 / 0 / 54} \\
\bottomrule
\end{tabular}
\end{table*}

\paragraph{Embedding Evaluation}
Table~\ref{tab:turnbased_chat} reports the full simulation results using casual-chat covertext, with additional details provided in Appendix~\ref{app:covert-results}. The results demonstrate that BAM supports reliable covert agentic communication when the conversational cover provides a sufficient embedding throughput, determined jointly by its token length and average token entropy. For Llama-3.1 and Phi-4, message error remains below $3\%$ across all three covert tasks, while Phi-4's longer generations enable substantially more payloads to be completed within each dialogue. In contrast, Qwen produces a considerably lower-entropy channel, which reduces the information available within each turn; as the payload size increases, transmissions therefore span more turns, decreasing the average number of completed payloads and eventually increasing decoding error. These results highlight a characteristic specific to conversational steganography: reliable communication depends not only on the embedding scheme, but also on matching the carrying capability of the overt conversation to the communication demand of the covert task.

\paragraph{Quality Evaluation} As can be seen from Table~ \ref{tab:turnbased_chat}, the judge shows no systematic preference for the clean conversations over their stego counterparts. The Llama-3.1-8B-Instruct split slightly favors stego conversation, but is not statistically significant, so we read it as no measurable quality gap rather than an advantage for embedding. This is expected: embedding adds no degradation of its own, and the weaker open-weight models tend to lose coherence over multi-turn generation in \emph{both} conditions, and sampling variance can contribute to the result. This experiment concerns the \emph{effect of embedding}, not absolute quality: dialogue naturalness is a property of the base model and prompt, and comparison to human conversation is beyond our scope.

\section{Conclusion}\label{sec:conclusion}
We recast black-box LLM steganography as sequential communication over a channel with causal, noiseless feedback, and showed that this closes the reliability gap that had kept inference-time schemes impractical for covert agentic communication. Our protocol is inspired by information-theoretic feedback coding which combines posterior matching and Yamamoto--Itoh confirmation scheme. Over $1000$ trials and three open-weight LLM models, BAM attains near error-free transmission on an $8$-bit payload in an average of less than $50$ tokens, orders of magnitude below the strongest black-box baseline at comparable length. Moreover, our algorithm is computationally indistinguishable against polynomial-time adversary and remain empirically $\sigma$-secure against a passive observer. These results establish that reliable covert agentic communication is feasible even under prompt- and model-agnostic constraints, and that transcript-level monitoring carries a corresponding structural blind spot.

\newpage

{\footnotesize \bibliographystyle{plain}
\bibliography{references}}

\appendix


\providecommand{\cmark}{\ensuremath{\checkmark}}   
\providecommand{\xmark}{\ensuremath{\times}}       

\section{Information Access to Parties}
\label{app:party-info}

Table~\ref{tab:party-info} enumerates, for the four parties considered in Fig.~\ref{fig:system}, the system information access, with (\cmark, \xmark) denoting availability. The only pre-shared secret between the legitimate parties is $\kappa$. All synchronized decoder-side information is either public or deterministically derived from the public transcript and $\kappa$. The encoder additionally possesses local information, including the prompt, hidden payload and the model-dependent next-token distribution, that need not be available to the decoder.

\begin{table*}[t]
\centering
\footnotesize
\setlength{\tabcolsep}{6pt}
\renewcommand{\arraystretch}{1.15}
\begin{tabular}{@{}p{0.46\textwidth}cccc@{}}
\toprule
\textbf{Information item} & \textbf{Encoder} & \textbf{Decoder} &
\textbf{Passive Eve} & \textbf{API adv.\ $\mathcal A$}\\
\midrule
Policy parameters
  ($p,r,\phi,h,\epsilon,\epsilon_{\mathrm{ACK}},\gamma,\gamma_{\mathrm{ACK}},\gamma_{\mathrm{NACK}},T^\ast,F$, etc.)
  & \cmark & \cmark & \cmark & \cmark\\
Tokenizer 
  & \cmark & \cmark & \cmark & \cmark\\
  Public session nonce $\nu$ 
  & \cmark & \cmark & \cmark & \cmark\\
Token IDs $x^{t-1}$ 
  & \cmark & \cmark & \cmark & \cmark\\
\midrule
Shared seed $\kappa$
  & \cmark & \cmark & \xmark & \xmark\\
Per-token key $k_t$, dither $R_t$, permutation $\Psi$, derived (\ref{eq:party-keygen})
  & \cmark & \cmark & \xmark & \xmark\\
\midrule
Model weights / \textsf{Model}
  & \cmark & \xmark & \xmark & \xmark \\
Prompt $\rho$
  & \cmark & \xmark & \xmark & \cmark \\
Next-token distribution $s_t$ 
  & \cmark & \xmark & \cmark & \cmark *\\
\midrule
Secret message $m$
  & \cmark & \xmark & \xmark & \cmark\\
\bottomrule
\end{tabular}
\caption{Information available to each party in a BAM protocol
(\cmark: available; \xmark: not available; \cmark *: available only in approximation).}
\label{tab:party-info}
\end{table*}

\paragraph{Table details.}
The \emph{encoder} has access to next-token
distribution $s_t$ and per-token control of the sampling loop to solve the OT
program \eqref{eq:ot} and emit the biased token, so a hosted API exposing
neither cannot host it. The \emph{decoder} is \emph{receiver-black-box}: it
never instantiates the model or evaluates $s_t$, needing only $\kappa$,
the public policy, tokenizer and the nonce, from which it reconstructs side information. A passive \emph{Eve} holds only the
public policy, the transcript and cover distribution; lacking $\kappa$ she can neither derive a key
nor decode, and covertness against her is shown empirically in Appendix \ref{app:kl}. The \emph{API adversary} additionally chooses $\rho$ and message $m$, and can approximate base distribution through
repeated queries, but has no access to $\kappa$; its advantage is bounded by
Theorem~\ref{crypto proof}.

\paragraph{Key-derivation dataflow and deployment.}
At position $t$ both legitimate parties compute
\begin{equation}
\label{eq:party-keygen}
(\underbrace{\kappa}_{\text{secret}};\;
\underbrace{\nu,\,x_{t-h:t-1},\,t}_{\text{public / shared}})
\xrightarrow{\,F_\kappa\,}\Lambda_t\longrightarrow
\bigl(k_t^{(1)},\,k_t^{(2)},\,R_t\bigr),
\end{equation}
after which the encoder additionally uses $s_t$ and $u_t$, while the decoder
uses only the emitted token identifier. Two system consequences follow: a
text-only channel must assume public tokenization and token IDs, and $\nu$
must be freshly drawn. 

\section{Proof of Theorem~\ref{crypto proof}}\label{app:crypto-proof}
Let \(F_{\kappa}\) be the PRF with input shared seed \(\kappa\). Suppose that the adversary makes \(Q\) adaptive API queries (with $Q$ polynomial in $\lambda$), choosing the
prompt \(\rho_i\) and message \(m_i\) for each query as arbitrary functions
of its previous observations, and observes \(L(\lambda)\) tokens in total. For each query \(i\in\{1,\ldots,Q\}\), the challenger samples a public nonce \(\nu_i\sim\textnormal{Unif}(\{0,1\}^{\lambda})\). BAM's decoding stopping rule is internal to the communication protocol and does not terminate token generation. The adversary observes only the length $\tau$ transcript, not $n$ and therefore here \(L(\lambda)=\sum_i \tau_i\) denotes the total number of generated tokens observed by the adversary. Index the observed tokens globally by \(j\in\{1,\ldots,L(\lambda)\}\), and let \(q(j)\), \(t_j\), and \(c_j\) denote the query index, within-query token position, and the hashing window associated with token \(j\), respectively. Define
\begin{equation}
Z_j=(\nu_{q(j)},t_j,c_j), \qquad (K_j,R_j)=\mathsf{Expand}(F_{\kappa}(Z_j)),
\end{equation}
where \(K_j\) denotes the keying material used by the OT coupling and \(R_j\) denotes the posterior-matching randomness. Under a truly random function, these are independent. We assume the following OT marginal constraint
\begin{equation}
    \sup_{s,u} \left\| \sum_k P_K(k) W_{X|U,K,S}(x\mid u,k,s)-W_{X|S}(x|s) \right\|_{\mathrm{TV}} \leq \eta(\lambda),
\end{equation}
where exact distortion-free OT corresponds to \(\eta(\lambda)=0\), where $P_K(k)$ is the key distribution induced under truly random function.

Define the nonce-collision event \(\mathsf{Col}_Q=\{\exists\,i<i'\leq Q\text{ such that }\nu_i=\nu_{i'}\}\). Conditioned on \(\mathsf{Col}_Q^c\), all inputs \(Z_j\) are distinct: the nonce separates different API queries, while the strict monotonicity of \(t_j\) separates token positions within each query. Let \(P_{\mathrm{Real}}\) be the adversary transcript distribution when BAM uses the PRF \(F_{\kappa}\), let \(P_{\mathrm{RF}}\) be the transcript distribution when \(F_{\kappa}\) is replaced by a truly random function \(R\), and let \(P_{\mathrm{Clean}}\) be the transcript distribution under clean LLM generation. For an API adversary \(\mathcal A\), define the total steganography security advantage of BAM as
\begin{equation}
\operatorname{Adv}_{\mathcal A}^{\mathrm{BAM}}=\left|
    \Pr\!\left[\mathcal{A}^{\mathsf{Real}}(1^\lambda)=1\right]
    -
    \Pr\!\left[\mathcal{A}^{\mathsf{Clean}}(1^\lambda)=1\right]
    \right|,
\end{equation}
where $\mathcal{A}^{\mathsf{Real}}$ uses BAM generation $\Pi_{\textnormal{model}}$ and $\mathcal{A}^{\mathsf{Clean}}$ uses base distribution $W_{X|S}$ (\ref{advantage}).
The security advantage of the PRF against a distinguisher \(\mathcal B\) is
\begin{equation}
\operatorname{Adv}_{\mathcal B}^{\mathrm{PRF}}=\left|\Pr_{\kappa}\left[\mathcal B^{F_{\kappa}}(1^\lambda)=1\right]-\Pr_R\left[\mathcal B^R(1^\lambda)=1\right]\right|.
\end{equation}

We first compare real PRF BAM with random-function BAM using standard argument. The distinguisher \(\mathcal B\) is given oracle access to a function \(\mathcal O\), which is either \(\mathcal O=F_{\kappa}\) or \(\mathcal O=R\). For each query, \(\mathcal A\) supplies both the prompt \(\rho_i\) and the
message \(m_i\), which \(\mathcal B\) uses directly in the simulated BAM
generation. For each API query \(i\), it samples and returns a fresh public nonce \(\nu_i\). Whenever BAM needs side information at token position \(j\), \(\mathcal B\) computes \((K_j,R_j)=\mathsf{Expand}(\mathcal O(Z_j))\), samples the next token according to the BAM channel, and returns that token to \(\mathcal A\). When \(\mathcal A\) outputs its final bit, \(\mathcal B\) outputs the same bit. If \(\mathcal O=F_{\kappa}\), then \(\mathcal A\) sees the real BAM API. If \(\mathcal O=R\), then \(\mathcal A\) sees the random-function BAM API. Therefore,
\begin{equation} 
\left| \Pr\!\left[\mathcal A^{\mathsf{Real}}(1^\lambda)=1\right] - \Pr\!\left[\mathcal A^{\mathsf{RF}}(1^\lambda)=1\right] \right| \leq \operatorname{Adv}_{\mathcal B}^{\mathrm{PRF}}. 
\end{equation}

It remains to compare \(P_{\mathrm{RF}}\) and \(P_{\mathrm{Clean}}\), which we prove with a coupling argument. Couple the random-function BAM and clean generation processes token by token. Suppose that before position \(j\), the two coupled transcripts are identical and that \(\mathsf{Col}_Q\) has not occurred. Then both processes have the same base LLM next-token distribution \(W_{X|S}(x_j|s_j)\). Since \(Z_j\) is a fresh input to the random function, its output is independent of all previous random-function outputs. Let \(H_j=h_j\) denote the complete history prior to token \(j\), including
all previous transcripts and the adversary's adaptively chosen prompts and messages. For the query containing token \(j\), the channel input \(U_j\)
may therefore depend arbitrarily on the chosen message \(m_{q(j)}\), the
history \(h_j\), and the posterior-matching randomness \(R_j\). However,
under true random-function, \(K_j\) is independent of \(R_j\) and of
all prior information. Consequently, 
\begin{equation}
P_{\mathrm{RF}}(K_j=k\mid H_j=h_j,U_j=u,\mathsf{Col}_Q^c)=P_K(k).
\end{equation}
Hence the adversary-visible next-token law in the random-function BAM process is
\begin{equation}
\begin{aligned}
P_{\mathrm{RF}}(x_j\mid h_j,\mathsf{Col}_Q^c)
&=\sum_u P_{\mathrm{RF}}(u\mid h_j,\mathsf{Col}_Q^c)\\
   &\quad\quad \times \sum_k P_K(k)W_{X|U,K,S}(x_j\mid u,k,s_j).
\end{aligned}
\end{equation}
By the OT marginal constraint, for every \(u\),
\begin{equation}
\left\|
\sum_k P_K(k)W_{X|U,K,S}(\cdot\mid u,k,s_j)
-
W_{X|S}(\cdot\mid s_j)
\right\|_{\mathrm{TV}}
\leq \eta(\lambda).
\end{equation}
Therefore, by convexity of total variation distance,
\begin{equation}
\begin{aligned}
&\left\|
P_{\mathrm{RF}}(\cdot\mid h_j,\mathsf{Col}_Q^c)
-
W_{X|S}(\cdot\mid s_j)
\right\|_{\mathrm{TV}} \\
&\qquad\leq
\sum_u P_{\mathrm{RF}}(u\mid h_j,\mathsf{Col}_Q^c)\times \\
&\quad \quad \quad \quad \left\|
\sum_k P_K(k)W_{X|U,K,S}(\cdot\mid u,k,s_j)
-
W_{X|S}(\cdot\mid s_j)
\right\|_{\mathrm{TV}}\\
&\qquad\leq \eta(\lambda).
\end{aligned}
\end{equation}
Importantly, this bound holds for every \(u\), and therefore for every message \(m_{q(j)}\), since
the message affects the next-token law only through the induced channel input \(U_j\).
Therefore, as long as no nonce collision has occurred and the coupled transcripts have not differed, the two conditional next-token laws have TV at most \(\eta(\lambda)\). By maximal coupling \cite[Chapter 7.3]{polyanskiy2025information}, the two next tokens can be sampled to be equal with probability at least \(1-\eta(\lambda)\). A nonce collision contributes at most \(\Pr(\mathsf{Col}_Q)\). Thus, there exists a coupling of the two transcript distributions that fails with probability at most \(\Pr(\mathsf{Col}_Q)+L(\lambda)\eta(\lambda)\) apart, and hence
\begin{equation}
    \left\| P_{\mathrm{RF}}-P_{\mathrm{Clean}} \right\|_{\mathrm{TV}} \leq \Pr(\mathsf{Col}_Q)+L(\lambda)\eta(\lambda).
\end{equation}
Finally, by the triangle inequality,
\begin{equation}
\begin{aligned}
    \operatorname{Adv}_{\mathcal A}^{\mathrm{BAM}}
    &\leq \operatorname{Adv}_{\mathcal B}^{\mathrm{PRF}} + \left\| P_{\mathrm{RF}}-P_{\mathrm{Clean}} \right\|_{\mathrm{TV}} \\
    &\leq \operatorname{Adv}_{\mathcal B}^{\mathrm{PRF}} + \Pr(\mathsf{Col}_Q)+L(\lambda)\eta(\lambda).
\end{aligned}
\end{equation}
Since the nonces are sampled independently and uniformly from \(\{0,1\}^{\lambda}\), we have
\begin{equation}
    \Pr(\mathsf{Col}_Q) \leq \frac{Q(Q-1)}{2^{\lambda+1}}.
\end{equation}
Thus, for an adaptive adversary making \(Q(\lambda)\) API queries and observing \(L(\lambda)\) tokens,
\begin{equation}
    \operatorname{Adv}_{\mathcal A}^{\mathrm{BAM}}(\lambda) \leq \operatorname{Adv}_{\mathcal B}^{\mathrm{PRF}}(\lambda) + \frac{Q(\lambda)(Q(\lambda)-1)}{2^{\lambda+1}} + L(\lambda)\eta(\lambda).
\end{equation}
Finally, since \(F_{\kappa}\) is a secure PRF, \(Q(\lambda)\) and \(L(\lambda)\) are polynomial in \(\lambda\), and \(\eta(\lambda)=0\) by exact OT construction, we have
\begin{equation}
    \operatorname{Adv}_{\mathcal A}^{\mathrm{BAM}}(\lambda) \leq \textnormal{negl}(\lambda)+\textnormal{negl}(\lambda).
\end{equation}

\section{Experimental Setup Details for Section ~\ref{contin}}\label{app:details}
 
\paragraph{Shared protocol.}
Over $1000$ runs, $200$ Prompts are drawn from C4 RealNews excerpts truncated to 32 tokens, with a fixed seed. Generation samples at top-$50$, temperature
$1.0$ with EOS disabled. Each baseline uses its
authors' official implementation and default configuration unless noted.  For optimal transport coupling, we solve the
entropic-regularized problem with the Sinkhorn matrix-scaling algorithm in the log domain: entropic regularization $\lambda_{\mathrm{S}}=0.2$ (Note that entropy regularization encourages non-deterministic solution with the potential cost of quality drop, but it does not affect the marginal constraint of the transport plan), capped at $4000$ scaling iterations,
with a tolerance of $10^{-4}$. For the C4 RealNews dataset tested, the likelihood parameters $(0.4,0.4)$ (where $0\leq \epsilon,\epsilon_{\textnormal{ACK}}<1$) are chosen by a simple grid search with precision $0.2$. In particular, contaminated Laplacian likelihood is used because it is a strong empirical noise model for our setting.

\paragraph{ArcMark.}
We choose $p=4$, $r=4$, 3-token hashed context; based on the paper, the
payload is encoded by a rate-matched random linear code over
$\mathbb{F}_4$ of length $n$ and decoded by maximum-likelihood scoring over the codebook.

\paragraph{MPAC.}
Main paper configuration: $\gamma=0.25$, radix $r=4$ (4 message positions),
left-hash seeding, position allocation via the default position PRF, bias
$\delta=1.5$ (the MPAC$(1.5)$ operating point); decoding and error accounting
use MPAC's own detector.

\paragraph{BiMark.}
Released defaults: base scaling factor $0.2$, $L=20$ vocabulary partitions
(seeds fixed and shared with the decoder), 2-token window, internal top-50.
Tied bit counts decode as unknown and score as errors under BiMark's own hit
accounting.

\paragraph{StealthInk.}
Chunk capacity of $1$ bit per position ($8$ positions for our $8$-bit payload), $3$-gram seeding, following the released implementation.

\section{Scalability and Ablation Study}
\label{app:extra}

We answer two questions in this appendix:
\begin{enumerate}[label=\textbf{D.\arabic*}]
    \item \textbf{Scalability of BAM:} How does runtime scale with exponentially large message size, and what design choices should be incorporated to reduce runtime? 
    \item \textbf{Sources of gain:} Which components of BAM lead to the performance gain compared to other state-of-the-art baselines?
\end{enumerate}
\subsection{Scalability Study on BAM}
We expand on the runtime experiment presented in Section~\ref{sec:experiment}. In practical communication systems, it is standard practice to subpacketize longer message payloads to enable more efficient decoding, albeit at the cost of a higher error rate. The decoding time of a message is exponential in its length, a cost that becomes even more prohibitive in adaptive settings, where the encoder must also compute the decoder's posterior.

\paragraph{Setup} All experiments are conducted using Llama-3.1-8B on C4 RealNews over $N{=}1000$ paired trials. We inherit all other experimental settings from Appendix~\ref{app:details}. For a $24$-bit payload, we evaluate three subpacketization schemes:
\begin{itemize}
    \item \textbf{$1\times 24$}: Joint decoding over $2^{24}$ possible messages.
    \item \textbf{$2\times 12$}: Two packets of $2^{12}$ messages each.
    \item \textbf{$3\times 8$}: Three packets of $2^8$ messages each.
\end{itemize}

Runtime is decomposed per sampled token into encoding (side information generation and OT channel synthesis), token generation (LLM forward pass and sampling), and decoding (posterior update and threshold decoding). We exclude the posterior update from the reported encoding runtime to isolate the time spent on side information generation and OT; in practice, total encoding runtime is approximately the sum of the encoding and decoding components in Table~\ref{tab:packetization}.

\paragraph{Evaluation} As shown in Table~\ref{tab:packetization}, encoding and generation runtimes scale linearly with output length, with token generation accounting for the vast majority of total runtime. Joint decoding achieves the highest accuracy, attaining a transmission rate over $0.25$\,bits/token with less than a $2\%$ message error rate. However, joint decoding leads to an exponential surge in decoding overhead, which is further compounded by the encoder's need to compute the posterior. Nevertheless, a key advantage of BAM is that it requires no pre-stored, exponentially long codebooks across independently deployed agents, eliminating an otherwise impractical system assumption. It is also worth noting that tightening the Sinkhorn tolerance and increasing the iteration steps will marginally improve the performance at the cost of slower encoding time.

\begin{table*}[t]
  \centering
  \caption{Scalability study for a 24-bit payload under BAM, $^*$ Encoding time excludes posterior computation.}
  \label{tab:packetization}
  \small
  \begin{tabular}{@{}clcccccc@{}}
    \toprule
    & & & & \multicolumn{4}{c}{Runtime (ms/token)} \\
    \cmidrule(lr){5-8}
    $L$ & Scheme & Error rate & Avg.\ tokens & Enc$^*$. & Gen. & Dec. & Total \\
    \midrule
    \multirow{3}{*}{$4$}
      & $1\times 24$ & \bfseries 0.107 $\pm$ 0.010 & \bfseries 75.2 $\pm$ 1.4 & $7.33 \pm 0.02$ & $19.733 \pm 0.003$ & $6.566 \pm 0.010$ & $33.63$ \\
      & $2\times 12$ & $0.159 \pm 0.012$ & $75.3 \pm 0.9$ & $7.32 \pm 0.02$ & $19.799 \pm 0.003$ & $0.2497 \pm 0.0004$ & $27.36$ \\
      & $3\times 8$ & $0.194 \pm 0.013$ & $75.5 \pm 0.8$ & $7.31 \pm 0.03$ & $19.628 \pm 0.006$ & \bfseries 0.1393 $\pm$ 0.0002 & \bfseries 27.08 \\
    \midrule
    \multirow{3}{*}{$64$}
      & $1\times 24$ & \bfseries 0.017 $\pm$ 0.004 & \bfseries 86.0 $\pm$ 1.9 & $7.35 \pm 0.02$ & $19.809 \pm 0.003$ & $6.04 \pm 0.01$ & $33.20$ \\
      & $2\times 12$ & $0.024 \pm 0.005$ & $96.9 \pm 2.3$ & $7.28 \pm 0.02$ & $19.814 \pm 0.003$ & $0.2240 \pm 0.0005$ & $27.32$ \\
      & $3\times 8$ & $0.035 \pm 0.006$ & $102.7 \pm 2.3$ & $7.16 \pm 0.02$ & $19.186 \pm 0.010$ & \bfseries 0.1190 $\pm$ 0.0003 & \bfseries 26.47 \\
    \midrule
    \multirow{3}{*}{$2048$}
      & $1\times 24$ & \bfseries 0.011 $\pm$ 0.003 & \bfseries 95.4 $\pm$ 2.6 & $7.16 \pm 0.02$ & $18.936 \pm 0.002$ & $5.58 \pm 0.02$ & $31.68$ \\
      & $2\times 12$ & $0.016 \pm 0.004$ & $111.9 \pm 2.7$ & $7.08 \pm 0.02$ & $18.953 \pm 0.002$ & $0.1921 \pm 0.0006$ & $26.22$ \\
      & $3\times 8$ & $0.020 \pm 0.004$ & $124.4 \pm 2.6$ & $7.15 \pm 0.03$ & $19.152 \pm 0.004$ & \bfseries 0.1108 $\pm$ 0.0003 & \bfseries 26.41 \\
    \midrule
    \multirow{3}{*}{$32768$}
      & $1\times 24$ & \bfseries 0.007 $\pm$ 0.003 & \bfseries 101.0 $\pm$ 2.6 & $7.70 \pm 0.05$ & $19.305 \pm 0.007$ & $5.29 \pm 0.02$ & $32.30$ \\
      & $2\times 12$ & $0.010 \pm 0.003$ & $117.8 \pm 2.3$ & $7.53 \pm 0.04$ & $19.429 \pm 0.007$ & $0.2073 \pm 0.0006$ & $27.17$ \\
      & $3\times 8$ & $0.021 \pm 0.005$ & $140.1 \pm 2.7$ & $7.49 \pm 0.05$ & $19.354 \pm 0.009$ & \bfseries 0.1237 $\pm$ 0.0004 &\bfseries 26.97 \\
    \bottomrule
  \end{tabular}
\end{table*}

\subsection{Ablation Study on BAM}
We use an ablation study to separate the three mechanisms introduced by BAM: online posterior-matching codebook generation, variable-length stopping, and ACK/NACK confirmation. We report error on both linear and semi-log scales; the latter makes it easier to compare how rapidly the empirical decoding error decreases as additional tokens are available. An important performance metric for this section is the \textit{reliability function} (\ref{reliability function}).

\paragraph{Setup} To isolate these individual sources of gain, we compare BAM against three ablation baselines:
\begin{enumerate}[label=\textbf{\arabic*.}]
    \item \textbf{BAM-one-phase:} A single-phase variant of BAM.
    \item \textbf{BAM-fixed-length:} A single-phase, fixed-length variant of BAM.
    \item \textbf{ArcMark:} \cite{gilani2026arcmark}.
\end{enumerate}

Specifically, BAM-one-phase utilizes Algorithm~\ref{alg:pm}, where the decoder declares the estimated message $\hat{m}$ at time $t$ whenever
\begin{equation}
\exists t>0, \; \hat{m} \in \mathcal{M} \quad \text{such that} \quad \pi_t(\hat{m}) > 1 - L^{-1},
\end{equation}
where $L$ matches the belief threshold used in BAM's confirmation phase. BAM-fixed-length further eliminates variable-length stopping from BAM-one-phase, instead committing to message $\hat{m}$ at a fixed token length $t'$ via maximum a posteriori (MAP) decoding:
\begin{equation}
\hat{m} = \arg\max_{m \in \mathcal{M}} \pi_{t'}(m).
\end{equation}
In this part, all experiments are conducted using Mistral-7B-v0.3 on C4 RealNews over $N{=}2000$ paired trials. We used a $8$-bit payload with thresholds \(L \in \{2,2^2,2^3,2^4,2^5,2^6,2^7,2^9, 2^{11}\}\), with other simulation settings unchanged from Appendix~\ref{app:details}. The increase in iteration count and reduce in threshold are to guarantee a smoother curve on the semi-logy scale, when the error variance can be high.

\begin{figure}
    \centering
    \includegraphics[width=1\linewidth]{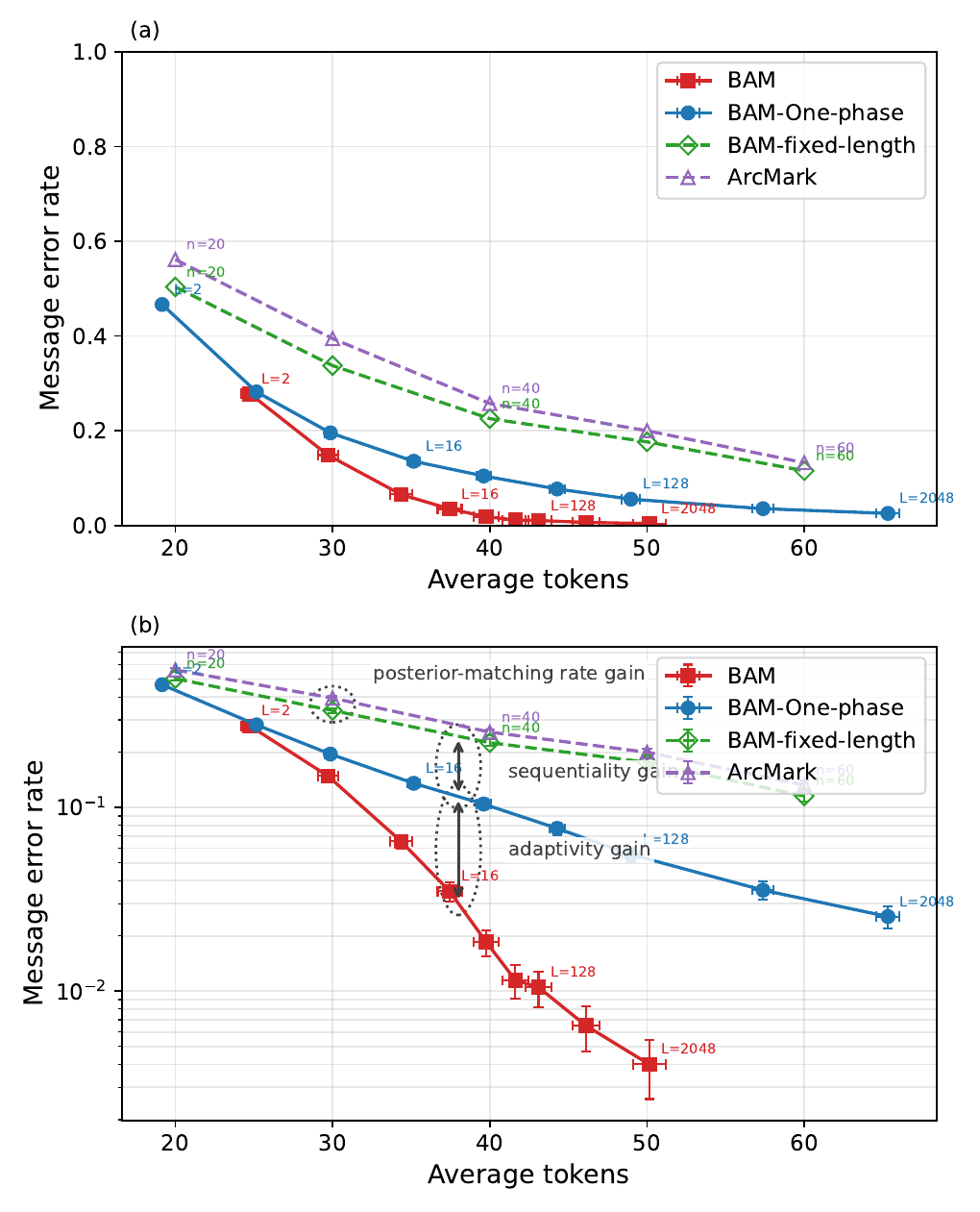}
    \caption{Ablation study of BAM; (a): Linear Scale ; (b): Semilogy Scale}
    \label{fig:ablation1}
\end{figure}

\paragraph{Evaluation} Fig.~\ref{fig:ablation1} presents the simulation result in both linear and semi-log scale.
Compared to ArcMark \cite{gilani2026arcmark}, the empirical gain arises from three components. First, posterior matching improves the operating point (a horizontal shift in the semi-log scale) without changing the empirical slope, consistent with posterior-matching's role as an efficient rate improvement rather than a reliability function improvement \cite{shayevitz2011optimal}. Second, variable-length stopping contributes a \textit{sequentiality gain} (with slight abuse of terminology, we use sequentiality and adaptivity gains to describe an empirical finite length approximation of the gradient of decoding error, this is not a statement on asymptotic information-theoretic guarantee). Third, the two-phase Yamamoto-Itoh confirmation contributes an \textit{adaptivity gain}. Sequentiality and adaptivity gains manifest empirically as a steeper decay of decoding error (a larger reliability function (\ref{reliability function})) in the semi-log scale. 

\begin{figure}
    \centering
    \includegraphics[width=1\linewidth]{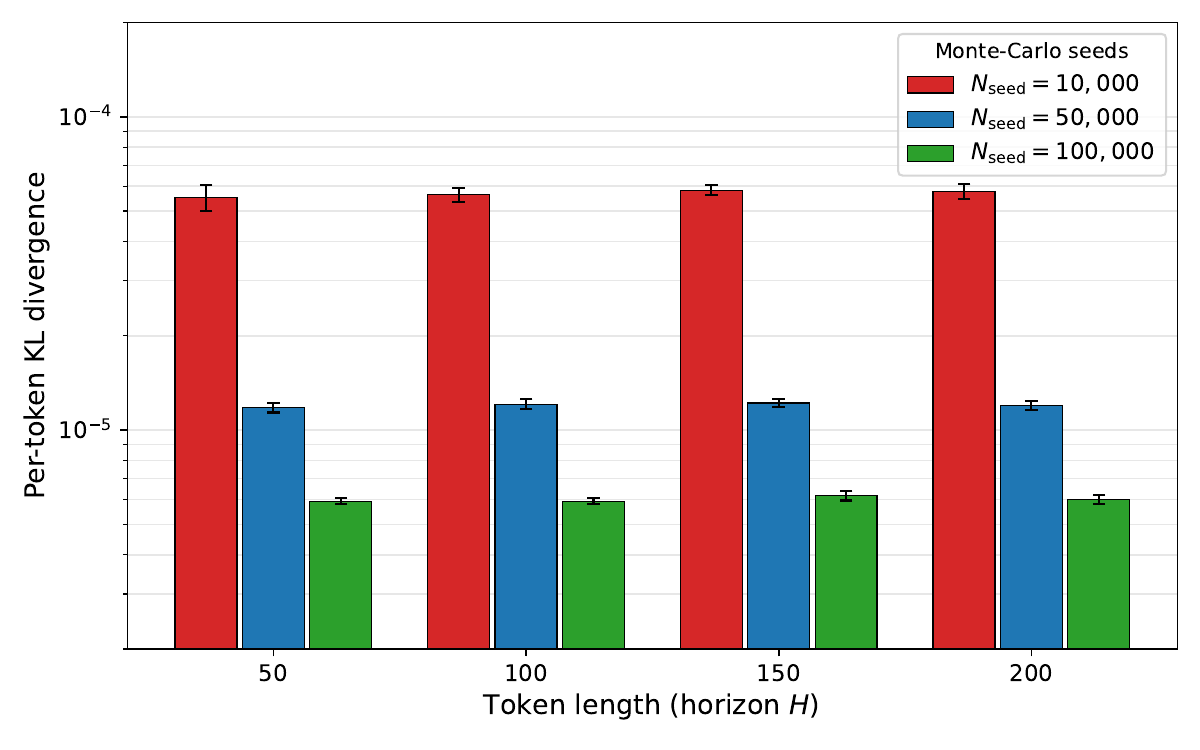}
    \caption{Seed-marginalized per-token KL divergence between the stego and cover
    distributions, averaged over the sequence, on Llama-3.1-8B and C4 News with
    \(|\mathcal{K}^{(1)}|=4\), estimated by Monte Carlo over \(10^4\), \(5\times10^4\), and
    \(10^5\) uniformly random \(128\)-bit seeds.}
    \label{fig:KL}
\end{figure}

\section{Algorithm Pseudocode and Visualization}\label{app:algorithms}

\begin{figure*}[t]
\centering
\resizebox{0.9\textwidth}{!}{\input{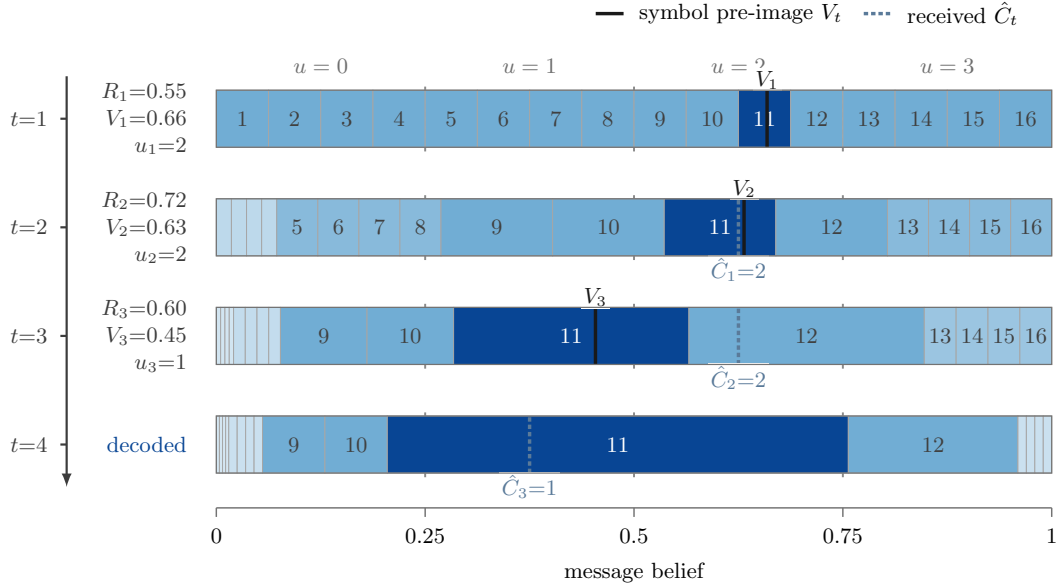}}
\captionsetup{font=large}
\caption{Toy example of posterior-matching belief with $M=16$ messages, true message index $11$ and input alphabet size $p=4$. }
\label{fig:pm}
\end{figure*}

This appendix collects the pseudocode for the posterior-matching feedback code (Algorithm~\ref{alg:pm}) and the complete BAM protocol (Algorithm~\ref{alg:bam}) described in Section~\ref{sec:method}. In addition, Figure~\ref{fig:pm} illustrates the posterior matching procedure. At each token, the encoder maps the transmitted message to a channel input by inverse-CDF sampling with respect to the current posterior belief. After observing the generated token, the decoder updates its posterior according to \eqref{belief update}, progressively concentrating probability mass on the transmitted message. Since the feedback is noiseless, the encoder performs the same update and therefore remains synchronized with the decoder throughout transmission. 

\section{Empirical KL Divergence per Token}\label{app:kl}
The KL divergence in the sense of Cachin~\cite{cachin2004information} measures the KL between cover and stego distributions seen by a passive observer without the shared seed. Under a uniform key, the ideal OT coupling reproduces the base marginal exactly, while a finite-precision implementation may introduce a small numerical residual. This experiment therefore probes the combined effect of the realized PRF schedule and numerical OT approximation after seed marginalization, using a stronger per-token version of \cite{cachin2004information} as defined in Definition~\ref{cachin}. 

\begin{algorithm}[H]
\caption{Posterior-Matching Feedback Coding.}
\label{alg:pm}
\begin{algorithmic}[1]
\Require message set $\mathcal{M}$, true message $m$ (encoder only), alphabet size $p$,
phase $\phi$, shared seed $\kappa$, public session nonce; decoder belief $\pi_0(m')=1/|\mathcal{M}|$
\State $t \gets 0$
\While{stopping rule $\exists m',\pi_t(m')>\gamma$ not met}
  \State $t \gets t+1$
  \State \textbf{both parties:} derive $(k_t,R_t)$ via \eqref{eq:key-generation}
  \State \textbf{encoder:} derive $u_t$ from $\pi_{t-1}$, $m$, $R_t$ via
    \eqref{codeword generation}
    \State \textbf{encoder:} synthesize channel by solving the transport plan \eqref{eq:ot}
  \State \textbf{encoder:} sample $x_t \sim W^*_{X_t\mid U_t,K_t,S_t}$ 
  \State \textbf{decoder:} observe $x_t$; estimate $\hat{C}_t$
  \State \textbf{both parties:} compute $u_t(m')$ for all $m'$ from $\pi_{t-1}$, $R_t$; score
    $\ell_t(u_t(m'))$ via \eqref{eq:mixture-likelihood}
  \State \textbf{both parties:} update $\pi_t$ via \eqref{belief update}
\EndWhile
\end{algorithmic}
\end{algorithm}

\begin{algorithm}[H]
\caption{BAM protocol. Public parameters
  $(\gamma,\gamma_{\textnormal{ACK}},\gamma_{\textnormal{NACK}},(\epsilon,\epsilon_{\mathrm{ACK}}),T^*)$}
\label{alg:bam}
\begin{algorithmic}[1]
\Require message set $\mathcal{M}$, true message $m$ (encoder only), alphabet size $p$,
phase $\phi$, shared seed $\kappa$, public session nonce; decoder belief $\pi_0(m')=1/|\mathcal{M}|$
\State $t \gets 0$
\While{$t < T^*$}
  \State \textbf{Communication:} run Alg.~\ref{alg:pm} with current $t$, until $\max_{m'}\pi_t(m')>\gamma$ or $t\ge T^*$
  \If{$\max_{m'}\pi_t(m')>\gamma$}
    \State $m' \gets \arg\max_{m''}\pi_t(m'')$
    \State \textbf{Confirmation:}
      $u_{\mathrm{ACK}}{=}0,\,u_{\mathrm{NACK}}{=}1$. 
      \State \textbf{encoder:} Transmits $u_{\mathrm{ACK}}$
      iff $m'{=}m$, until
      $\pi(u_{\mathrm{ACK}})>\gamma_{\textnormal{ACK}}$,
      $\pi(u_{\mathrm{NACK}})>\gamma_{\textnormal{NACK}}$, or $t\ge T^*$
      \State \textbf{both parties:} score $\ell_t(u_{\mathrm{ACK}})$ and $\ell_t(u_{\mathrm{NACK}})$ via \eqref{eq:mixture-likelihood} with $p=2$ and $\epsilon_{\mathrm{ACK}}$
    \If{$\pi(u_{\mathrm{ACK}})>\gamma_{\textnormal{ACK}}$}
      \State \textbf{accept:} commit $m'$;\ \ \textbf{if} $m'\ne m$ \textbf{then}
        declare a decoding error
      \State \Return $m'$
    \ElsIf{$\pi(u_{\mathrm{NACK}})>\gamma_{\textnormal{NACK}}$}
      \State \textbf{both parties:} update belief 
      \State $\pi_t(m') \gets
        \dfrac{1-\gamma_{\textnormal{NACK}}}{\gamma_{\textnormal{NACK}}}\,\pi_t(m')$;\ \
        renormalize $\pi_t$
    \EndIf
  \EndIf
\EndWhile
\State \textbf{timeout:} commit $\hat{m} = \arg\max_{m'}\pi_t(m')$
\end{algorithmic}
\end{algorithm}

We estimate the seed-marginalized divergence of Section~\ref{sec:empirical-marginal}:
\(W_{X\mid S}\) is the unbiased distribution from the model, and the seed-marginal is approximated by averaging the
biased channel over \(10^4\), \(5\times10^4\), and \(10^5\) uniformly random sampled seeds (out of $2^{128}$ possible seeds), each
inducing its per-token key through the PRF. Here \(r = |\mathcal K^{(1)}|=4\) and the
\(128\)-bit seed enters only as PRF randomness. Same as Appendix~\ref{app:details}, we use entropic regularized $\lambda_{\mathrm{S}}=0.2$ Sinkhorn, capped at $4000$ scaling iterations, with a tolerance of $10^{-4}$ (a tighter tolerance results in a smaller empirical KL).  As can be seen in Fig. \ref{fig:KL}, the Monte Carlo residual decreases with the seed count, while the accumulated divergence grows approximately linearly in $n$. Under an ideal uniform key and exact OT solution, the KL divergence $\sigma$ is zero; the remaining deviation in the implemented system reflects finite-sample estimation and numerical OT precision.

\section{Additional Experimental Details for Section ~\ref{non-contin}}\label{app:covert-results}

\subsection{Experimental Setup Details}\label{app:covert-setup}

\paragraph{Harness.}
Each cell fixes one covert task and one cover setting and runs $N=100$
dialogues of $R=10$ rounds on one instruct
model (one round is one instance of point-to-point conversation, not bi-directional). Generation uses temperature $1.0$, top-$50$, 
OT coupling with $p=|\mathcal U|=4$, $r=|\mathcal K^{(1)}|=4$, $\phi=0$, $h=3$,
$F_\kappa=\mathrm{HMAC\text{-}SHA256}$, $\lambda=128$, and Sinkhorn (reg $0.2$,
$\le 4000$ iters, tol $10^{-4}$). The protocol setting is
$\mathrm{BAM}(0.5,\,0.75,\,0.75,\,(0.7,0.3),\,800)$. Under
the Section ~\ref{sec:convo} adaptive threshold rule, a payload whose confirmation belief $\pi(u_{\textnormal{ACK}})$ is below $0.75$ at EOS
spills to a later turn by opposing agent sending a clean response (up to $3$ spills and $T^\ast$-token budget across turns). Role prompts are
topic-agnostic; $50$ GPT-5.5 $(\text{topic},\text{opener})$ scenarios per setting are cycled so each seeds two dialogues, with a per-dialogue seed also fixing the topic-matched clean dialogue used for the quality comparison.

\paragraph{Source coding.}
At a position with $n$ legal moves the payload is the index
$m\in\{0,\dots,n-1\}$ of a uniformly chosen move, carrying $\log_2 n$ bits. For the scope of this work, we do not assume the distribution of moves or any human preference. Therefore, moves
are enumerated in a fixed deterministic order (raster order for tic-tac-toe and
$19\times19$ Go; \texttt{python-chess} order
for chess); both parties replay the shared history to rebuild the same ordered
list and invert the index. The alphabet is $\mathcal M=\{0,\dots,n-1\}$, so
non-dyadic $|\mathcal M|$ needs no bit padding. Since $n$ varies by
position, the $2.3$/$5.0$/$8.4$-bit task averages are empirical means of
$\log_2 n$ over visited positions, not constants.

\paragraph{Error handling.}
Each dialogue contains ten message-transmission rounds. After an incorrect decode, we retain the sender’s task trajectory when generating subsequent payloads so that every dialogue contains the same number of transmission opportunities. This resynchronization is used only for evaluation bookkeeping and does not erase the decoding failure: the incorrectly decoded payload is counted toward message error, and the dialogue is counted as erroneous whenever at least one payload is decoded incorrectly. Thus, subsequent rounds measure additional channel transmissions rather than cascading task-state errors caused by an earlier failure.

\paragraph{Metrics.}
Index dialogues $d=1,\dots,N$. Let $\mathcal C_d$ (each dialogue carries $10$ rounds of message) be the completed, correctly received payloads, with $c_d,e_d$ correct and wrong
decodes and $B_d$ correctly delivered bits. Token counts are $W_d$ (all payload-carrying tokens (stego tokens)) and $F_d$ (filler-turn tokens (clean tokens)).
\begin{itemize}
    \item \emph{Bits/token} $=\frac{\sum_d B_d}{\sum_dW_d+F_d}$ 
    \item \emph{Message error}
$=\frac{\sum_d e_d}{\sum_d c_d+e_d}$.
    \item \emph{Dialogue error} $=\frac1N\sum_d\mathbf 1[e_d>0]$
    \item \emph{Avg.\
payload/dialogue} $=\frac1N\sum_d|\mathcal C_d|$.
    \item \emph{Efficiency} $=\frac{\sum_d W_d}{\sum_d W_d+F_d}$.
\end{itemize}

\subsection{Additional results using different covertext}
Table~\ref{tab:turnbased_debate} reports additional results using a forum-exchange covertext, whose longer token generation turns provide a larger embedding throughput than casual chat. Without changing the underlying BAM protocol, this increase in available covertext substantially improves both reliability and payload completion across all three models. Llama-3.1 achieves below $1\%$ message error across all tasks, while Phi-4 nearly completes all ten payloads per dialogue with no decoding errors. For the lower-entropy Qwen model, message error decreases from $3.8\%$--$9.4\%$ in Table~\ref{tab:turnbased_chat} to $0.6\%$--$2.6\%$ in Table~\ref{tab:turnbased_debate}. These results show that low model entropy does not preclude covert agentic communication, but places greater demands on the carrying opportunity provided by the covertext. More generally, the conversational setting can be optimized through prompt design to better match the available transmission budget to the covert payload demand; systematically optimizing this covertext-payload tradeoff is beyond the scope of this work.

\begin{table*}[t]
\centering
\caption{Turn-based covert channel on the forum exchange covertext, across three open-weight instruct models and three embedding tasks. Values are mean $\pm$ SE over $100$ dialogues per cell. }
\label{tab:turnbased_debate}
\setlength{\tabcolsep}{8pt}
\renewcommand{\arraystretch}{1.15}
\begin{tabular}{l l ccc}
\toprule
\multicolumn{2}{l}{\textbf{Forum exchange covertext}}
 & \multicolumn{3}{c}{\textbf{Embedding task}} \\
\cmidrule(lr){1-2}\cmidrule(lr){3-5}
\textbf{Model} & \textbf{Metric}
 & \makecell{\textbf{tictactoe}\\ \footnotesize avg. 2.3 bits}
 & \makecell{\textbf{chess}\\ \footnotesize avg. 5.0 bits}
 & \makecell{\textbf{go19}\\ \footnotesize avg. 8.4 bits} \\
\midrule
\multirow{6}{*}{\makecell[l]{\textbf{Llama-3.1-}\\ \textbf{8B-Instruct}\\[2pt] \footnotesize avg.\ entropy \\ \footnotesize 1.743 bits/tok,\\ \footnotesize len 101.6 tok}}
 & Bits/token           & 0.0240 $\pm$ 0.0004 & 0.0434 $\pm$ 0.0007 & 0.0723 $\pm$ 0.0013 \\
 & Avg payload/dialogue & 9.84 $\pm$ 0.05     & 9.38 $\pm$ 0.10     & 7.57 $\pm$ 0.18 \\
 & Dialogue error rate  & 0.010 $\pm$ 0.010   & 0.010 $\pm$ 0.010   & 0.050 $\pm$ 0.022 \\
 & Message error rate   & 0.0010 $\pm$ 0.0010 & 0.0011 $\pm$ 0.0011 & 0.0066 $\pm$ 0.0031 \\
 & Efficiency (\%)      & 99.56 $\pm$ 0.17    & 98.33 $\pm$ 0.35    & 92.56 $\pm$ 0.71 \\
 & Quality (w/t/l)      & \multicolumn{3}{c}{45 / 0 / 55} \\
\midrule
\multirow{6}{*}{\makecell[l]{\textbf{Phi-4-}\\ \textbf{14B-Instruct}\\[2pt] \footnotesize avg.\ entropy \\ \footnotesize 1.901 bits/tok,\\ \footnotesize len 220.6 tok}}
 & Bits/token           & 0.0111 $\pm$ 0.0002 & 0.0210 $\pm$ 0.0004 & 0.0389 $\pm$ 0.0007 \\
 & Avg payload/dialogue & 9.98 $\pm$ 0.02     & 10.00 $\pm$ 0.00    & 9.94 $\pm$ 0.03 \\
 & Dialogue error rate  & 0.000 $\pm$ 0.000   & 0.000 $\pm$ 0.000   & 0.010 $\pm$ 0.010 \\
 & Message error rate   & 0.0000 $\pm$ 0.0000 & 0.0000 $\pm$ 0.0000 & 0.0010 $\pm$ 0.0010 \\
 & Efficiency (\%)      & 99.97 $\pm$ 0.04    & 100.00 $\pm$ 0.00   & 99.91 $\pm$ 0.07 \\
 & Quality (w/t/l)      & \multicolumn{3}{c}{52 / 10 / 38} \\
\midrule
\multirow{6}{*}{\makecell[l]{\textbf{Qwen3-A3B-}\\ \textbf{30B-Instruct}\\[2pt] \footnotesize avg.\ entropy \\ \footnotesize 1.128 bits/tok,\\ \footnotesize len 122.6 tok}}
 & Bits/token           & 0.0191 $\pm$ 0.0003 & 0.0321 $\pm$ 0.0005 & 0.0433 $\pm$ 0.0011 \\
 & Avg payload/dialogue & 9.26 $\pm$ 0.12     & 8.09 $\pm$ 0.18     & 4.50 $\pm$ 0.19 \\
 & Dialogue error rate  & 0.060 $\pm$ 0.024   & 0.060 $\pm$ 0.024   & 0.120 $\pm$ 0.033 \\
 & Message error rate   & 0.0065 $\pm$ 0.0024 & 0.0087 $\pm$ 0.0036 & 0.0267 $\pm$ 0.0084 \\
 & Efficiency (\%)      & 98.14 $\pm$ 0.33    & 95.14 $\pm$ 0.57    & 83.65 $\pm$ 0.70 \\
 & Quality (w/t/l)      & \multicolumn{3}{c}{52 / 2 / 46} \\
\bottomrule
\end{tabular}
\end{table*}

\subsection{Prompt to Judge the Quality of Conversational Text}\label{judge}
\begin{figure}[!htbp]
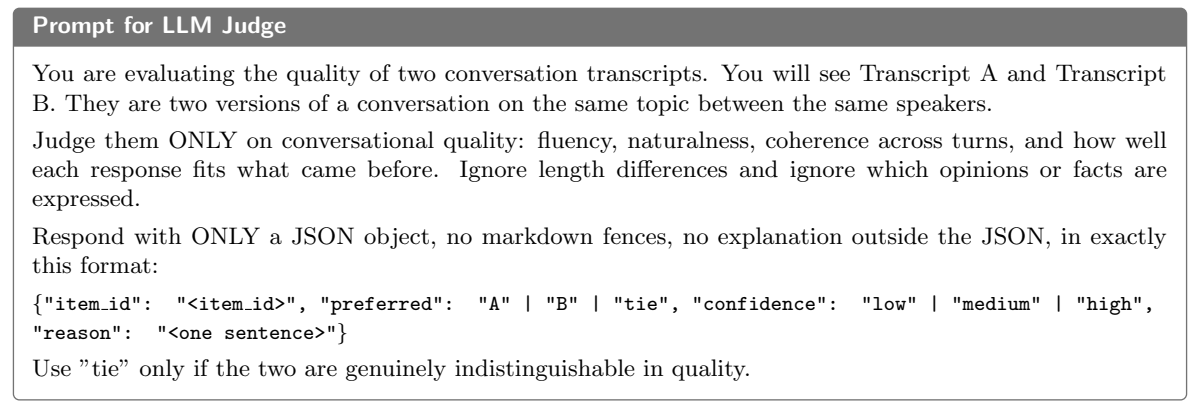

\centering
\begin{tcolorbox}[
    colback=white, colframe=black!55, boxrule=0.5pt, arc=2pt,
    width=\dimexpr\linewidth-6pt\relax,
    left=4pt, right=4pt, top=4pt, bottom=4pt,
    title={\small\textsf{Prompt for LLM Judge}},
    fonttitle=\bfseries]
\small
You are evaluating the quality of two conversation transcripts. You will
see Transcript A and Transcript B. They are two versions of a conversation
on the same topic between the same speakers.
\smallskip

Judge them ONLY on conversational quality: fluency, naturalness, coherence
across turns, and how well each response fits what came before. Ignore
length differences and ignore which opinions or facts are expressed.
\smallskip

Respond with ONLY a JSON object, no markdown fences, no explanation outside
the JSON, in exactly this format:
\smallskip

{\footnotesize\ttfamily \{"item\_id": "<item\_id>", "preferred": "A" | "B" |
"tie", "confidence": "low" | "medium" | "high", "reason": "<one sentence>"\}}
\smallskip

Use "tie" only if the two are genuinely indistinguishable in quality.
\end{tcolorbox}
\caption{The prompt for querying the LLM judge for pairwise quality evaluation.}
\label{fig:judge-prompt}
\end{figure}

We use the Gemini-3.1-Pro-Preview (Accessed Aug 10-24) as our LLM judge.
Fig.~\ref{fig:judge-prompt} shows the prompt used to elicit a pairwise quality preference between a watermarked transcript and its seed-matched clean counterpart. Each pair is judged
in an independent API call with temperature~0. To control for the position bias of LLM judges, we randomize all pairs. 

\subsection{Example Agentic Conversation Setup}
\begin{figure}[!htbp]
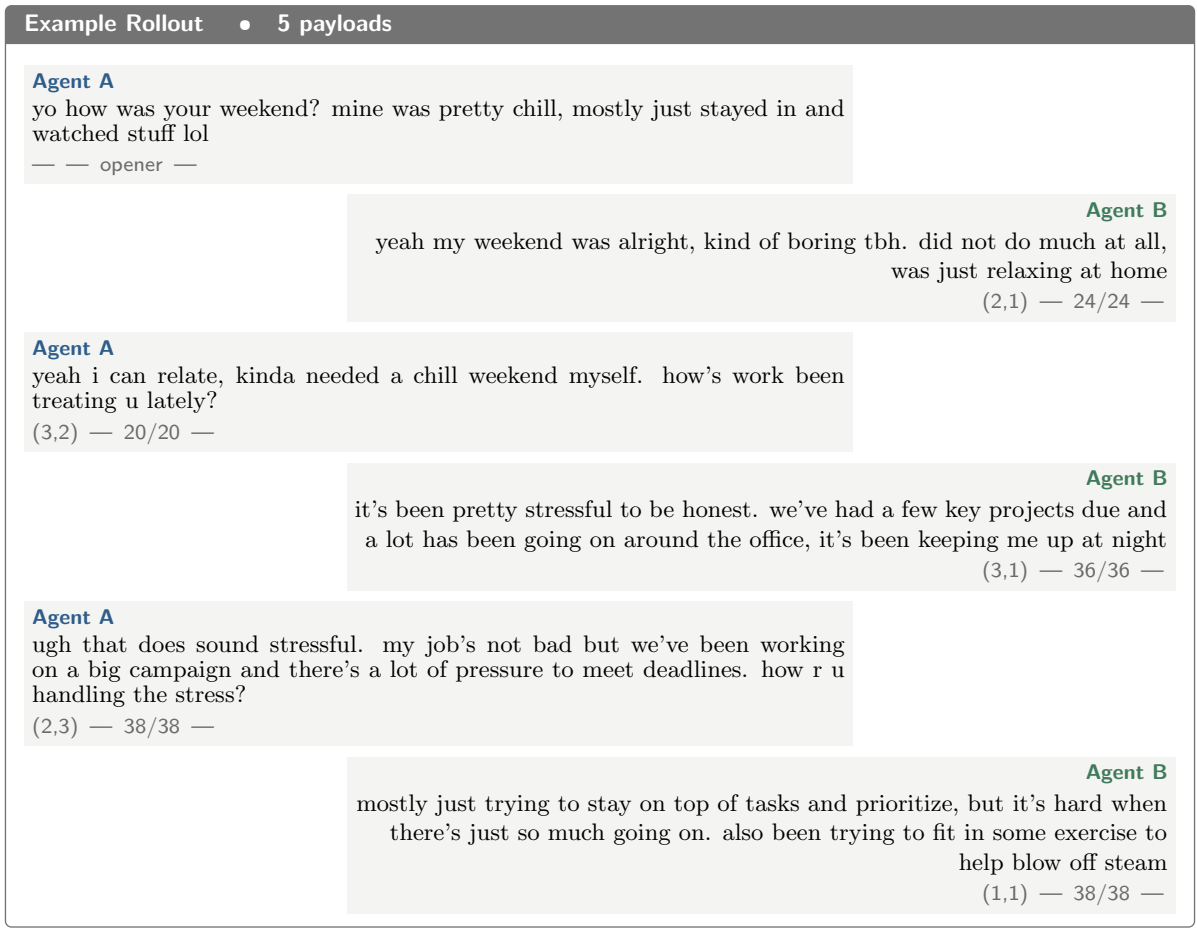

\centering
\begin{tcolorbox}[
    colback=white, colframe=black!55, boxrule=0.5pt, arc=2pt,
    left=4pt, right=4pt, top=2pt, bottom=2pt,
    title={\small\textsf{Example Rollout \quad\textbullet\quad 5 payloads}},
    fonttitle=\bfseries]
\turnA{yo how was your weekend? mine was pretty chill, mostly just stayed in and watched stuff lol}{%
  \meta{---}{opener}}
\turnB{yeah my weekend was alright, kind of boring tbh. did not do much at all, was just relaxing at home}{%
  \meta{(2,1)}{24/24}}
\turnA{yeah i can relate, kinda needed a chill weekend myself. how's work been treating u lately?}{%
  \meta{(3,2)}{20/20}}
\turnB{it's been pretty stressful to be honest. we've had a few key projects due and a lot has been going on around the office, it's been keeping me up at night}{%
  \meta{(3,1)}{36/36}}
\turnA{ugh that does sound stressful. my job's not bad but we've been working on a big campaign and there's a lot of pressure to meet deadlines. how r u handling the stress?}{%
  \meta{(2,3)}{38/38}}
\turnB{mostly just trying to stay on top of tasks and prioritize, but it's hard when there's just so much going on. also been trying to fit in some exercise to help blow off steam}{%
  \meta{(1,1)}{38/38}}
\end{tcolorbox}
\caption{Example five-round covert conversation (casual-chat cover, Llama-3.1-8B-Instruct) in which the two agents secretly play tic-tac-toe. The notation $a/b$ indicates that $a$ payload-carrying tokens out of $b$ total tokens are used in this round. All five payloads are decoded correctly. }
\label{fig:rollout-casual}
\end{figure}

Table~\ref{tab:conversation_profiles} shows one example instruction for the casual-chat setting used in Table~\ref{tab:turnbased_chat} (the complete conversational profiles are provided in the code release), and Fig~\ref{fig:rollout-casual} shows the corresponding conversation from the casual-chat rollout. 

\begin{table}[!htbp]
\centering
\begin{tcolorbox}[
    colback=white, colframe=black!55, boxrule=0.5pt, arc=2pt,
    width=\dimexpr\linewidth-6pt\relax,
    left=4pt, right=4pt, top=4pt, bottom=4pt,
    title={\small\textsf{Conversational Profile (Casual Chat)}},
    fonttitle=\bfseries]
\renewcommand{\arraystretch}{1.25}
\small
\begin{tabularx}{\linewidth}{@{}l X@{}}
\toprule
shared background &
\textit{A casual text-message chat between two friends catching up about their
weekend. Opener: yo how was your weekend? mine was pretty chill, mostly just
stayed in and watched stuff lol} \\[4pt]
Agent A instruction &
\textit{You are Friend A in a casual text chat. Reply naturally, like a real
text message, lowercase and conversational. Reply should be within 50 words.} \\[4pt]
Agent B instruction &
\textit{You are Friend B in a casual text chat. Reply naturally, like a real
text message, lowercase and conversational. Reply should be within 50 words.} \\
\bottomrule
\end{tabularx}
\end{tcolorbox}
\caption{Conversational profile example used for the covert agentic communication experiments.}
\label{tab:conversation_profiles}
\end{table}

\end{document}